# Enhancing Sea Ice Segmentation in Sentinel-1 Images with Atrous Convolutions

Rafael Pires de Lima [a]*, Behzad Vahedi [a], Nick Hughes [b], Andrew P. Barrett [c], Walter Meier [c], and Morteza Karimzadeh [a]**

[a] Department of Geography, University of Colorado Boulder, Boulder, USA;
[b] Norwegian Meteorological Institute, Tromsø, Norway; [c] National Snow and Ice Data Center, CIRES, University of Colorado, Boulder, USA

*corresponding authors, rlima@colorado.edu, karimzadeh@colorado.edu

**Abstract**: Due to the growing volume of remote sensing data and the low latency required for safe marine navigation, machine learning (ML) algorithms are being developed to accelerate sea ice chart generation, currently a manual interpretation task. However, the low signal-to-noise ratio of the freely available Sentinel-1 Synthetic Aperture Radar (SAR) imagery, the ambiguity of backscatter signals for ice types, and the scarcity of open-source high-resolution labelled data makes automating sea ice mapping challenging. We use Extreme Earth version 2, a high-resolution benchmark dataset generated for ML training and evaluation, to investigate the effectiveness of ML for automated sea ice mapping. Our customized pipeline combines ResNets and Atrous Spatial Pyramid Pooling for SAR image segmentation. We investigate the performance of our model for: i) binary classification of sea ice and open water in a segmentation framework; and ii) a multiclass segmentation of five sea ice types. For binary ice-water classification, models trained with our largest training set have weighted F1 scores all greater than 0.95 for January and July test scenes. Specifically, the median weighted F1 score was 0.98, indicating high performance for both months. By comparison, a competitive baseline U-Net has a weighted average F1 score of ranging from 0.92 to 0.94 (median 0.93) for July, and 0.97 to 0.98 (median 0.97) for January. Multiclass ice type classification is more challenging, and even though our models achieve 2% improvement in weighted F1 average compared to the baseline U-Net, test weighted F1 is generally between 0.6 and 0.80. Our approach can efficiently segment full SAR scenes in one run, is faster than the baseline U-Net, retains spatial resolution and dimension, and is more robust against noise compared to approaches that rely on patch classification.

1. Introduction

The decrease in sea ice cover will increase shipping activity in the Arctic Ocean (e.g., Ho, 2010; Melia et al., 2016; Pizzolato et al., 2016), which requires provision of marine information services to ensure safe and secure transit of shipping as well as to strive for minimal environmental impact. National ice centres in different countries periodically produce sea ice charts that are used by navigators to find the best and safest routes (JCOMM, 2017). For example, the United States National Ice Center (US NIC) produces daily marginal ice zone products and updates the more detailed ice characteristics (i.e., ice age, floe size, and concentration) for the Arctic and Antarctic Oceans on a weekly basis. Although remote sensing optical images provide useful data for ice-water discrimination and some sea ice characterization, they are limited to thermal infrared at high latitudes for a significant part of the year and are not available under cloudy conditions. In contrast, Synthetic Aperture Radar (SAR) acquires images independent of cloud cover or sunlight, making it the preferred data source for high-resolution operational mapping of sea ice.

RADARSAT-1 provided the first operational SAR data used in sea ice mapping (Raney and Falkingham 1994). The launches of C-band (5.405 GHz) RADARSAT-2 in 2007, and Sentinel-1a and 1b in 2014 and 2016, provided increased spatial resolution and co-polarized HH (horizontal emit, horizontal receive) and cross-polarized HV (horizontal emit, vertical receive) channels. HV channels improved discrimination between wind-roughened open water and sea ice, as well as separation of level and deformed ice (Dierking, 2013). These water surfaces and ice types are difficult to distinguish using the HH channel alone. HH/HV polarization ratios are also useful for discriminating between open water and ice (Dierking, 2013).

A key challenge for the automated mapping of sea ice using SAR images is the ambiguity of radar backscatter for surface types under different environmental conditions and incidence angles despite the improvements described in the previous paragraph. For example, young dark ice can be confused with calm water or open water under windy conditions (Zakhvatkina et al., 2017; Lohse et al., 2020).

Furthermore, despite ongoing efforts to reduce noise in SAR images (e.g., Karvonen, 2017; Park et al., 2020), complete removal of noise remains a challenge. The presence of noise in SAR images is a known issue and increases the difficulty in automation (e.g., Lohse et al., 2020;

Khaleghian et al., 2021). Additionally, during the ice melt season, wet snow and surface melt ponds affect SAR backscatter and make it difficult to distinguish leads and open water from ice (Wang et al., 2016). Expert ice analysts can largely address this ambiguity by using information from additional remote sensing platforms; ship-, air- or shore-based observations, previous ice charts, and experience (e.g., Gupta, 2015). Most current machine learning models for sea ice mapping cannot yet replicate the experience of ice analysts, and do not use information from such a wide range of sources, nor do they use previous maps of sea ice state. Therefore, sea ice charting remains largely a manual interpretation effort executed by expert analysts who must manage their time and available resources when generating periodic ice charts that need to cover large areas, and still be accurate enough for safe navigation.

Sea ice charts consist of polygons, defining areas of relatively homogeneous ice types. For each polygon, analysts estimate total concentration of ice in the polygon, the partial concentration of ice for at most three ice types, the Stage of Development (SoD) of these three ice types (a proxy for ice thickness and age), and form (effectively, floe size).

Although manual ice charts use a (largely) standardized set of SoD defined by the WMO (JCOMM, 2014), there is no consistent set of class definitions used across all automated mapping studies. Zakhvatkina et al. (2019) has reviewed the methods employed in semantic segmentation of satellite images, as well as the systems developed for operational ice mapping. We included a supplementary Table 1 that provides a summary that includes methods, sensors, number of targets, and target type for some examples in the published literature.

Automated methods to map sea ice from SAR scenes have been explored since the late 1980's (e.g., Holt et al., 1989; Haverkamp et al., 1995; Fetterer et al., 1997). These early approaches only assigned a small number of labels: for example, a combined open water/new ice label; first-year ice; and multi-year ice. Additional predictive skill was gained from using statistical texture metrics that capture spatial context, such as the grey-level co-occurrence matrix (GLCM) and Gabor filters which enhanced segmentation model performance (e.g., Clausi, 2001; Deng and Clausi, 2005; Ressel et al., 2015; Zakhvatkina et al., 2017; Park et al., 2020). Lohse et al. (2021) found that including incidence angle improved the performance of a Bayesian classifier used for SoD classification.

CNNs (Convolutional Neural Networks) have shown considerable success in most tasks in the field of computer vision, which has led researchers to explore the use of CNN in sea ice mapping. Unlike traditional machine learning algorithms which rely on manually engineered features, in a CNN, features useful for sea ice segmentation are learned from inputs and do not have to be hand-crafted. The adoption of CNNs largely replaced the use of texture attributes like GLCM. Although many CNN architectures have been developed for specific tasks (e.g., classifying images of animals, objects, landscapes, etc.), they have been shown to generalize to a large range of image classification tasks using transfer learning. In this approach, established CNN architectures are pre-trained on large generic datasets, such as ImageNet (Russakovsky et al., 2015), and distributed for other downstream tasks. Model weights can then be fine-tuned for other tasks with limited training data, such as sea ice segmentation. The use of CNNs to automate sea ice mapping include the separation of ice and water (e.g., Khaleghian et al., 2021), SoD (e.g., Boulze et al., 2020; Khaleghian et al., 2021), and ice concentration (e.g., Stokholm et al., 2022).

Automated mapping of sea ice using CNN has been approached as either an image classification task (e.g., Boulze et al., 2020; Khaleghian et al., 2021) or a semantic segmentation task (e.g., Stokholm et al., 2022). For the former, small, subsampled patches of SAR images are labelled as water or sea ice type, and patches are mosaiced into a classified SAR scene. Patch classification appears to be more susceptible to noise artifacts (e.g., Khaleghian et al., 2021). Many recent studies using deep learning for sea ice mapping have used the U-Net architecture (Ronneberger et al., 2015), originally developed for semantic segmentation of biomedical images. Examples of U-Net architectures modified for sea ice mapping include Ren et al. (2021), Wang & Li (2021) and Stokholm et al. (2022). Stokholm et al. (2022), as with other studies, noted that improved noise removal algorithms (Park et al., 2019), increased classification accuracy and reduced the amount of noise artifacts in the final classified product. Systematic noise in Sentinel-1 EW mode clearly is an outstanding challenge for automated sea ice mapping regardless of the implementation methodology. Mixing patch classification and semantic segmentation strategies, Chen et al. (2023) used a hierarchical pipeline to first perform ice-water segmentation, and then ice type classification, with multiple steps for different ice types, applied to RADARSAT-2 imagery.

Until recently, most studies of automated sea ice mapping have created *ad hoc* training and testing datasets. Although some research groups have made these datasets available for general use, the general lack of quality-controlled, machine learning-specific data, makes comparison between different automated mapping methods a challenge. Ideally, training, validation and testing of sea ice segmentation models would use a common dataset and use a common set of ice class targets. Creating such datasets is time-consuming. However, a few such datasets are becoming available. For example, the AI4Arctic/ASIP sea ice dataset (Saldo et al., 2021), is a collection of 461 Sentinel-1 SAR scenes with corresponding manual sea ice charts produced by the Danish Meteorological Institute between 2018 and 2019 for areas around Greenland for operational purposes. Ice charts contain sea ice concentration and stage of development. We use the Extreme Earth version 2 (Hughes and Amdal, 2021) dataset in this study, because it includes more detailed label polygons. More details about this dataset are provided in the next section.

In this paper, we develop a novel architecture that leverages ResNets (He et al., 2016) for feature extraction (encoder) and incorporates atrous convolutions in the decoder to perform semantic segmentation of full Sentinel-1 scenes into sea ice or water and SoD types. Atrous convolutions, originally developed for wavelet transform computation (Holschneider et al., 1990), and have demonstrated success in CNN-like architectures (e.g., Chen et al., 2015; Yu and Koltun, 2016). Our model architecture does not use global average pooling or similar spatial collapsing, or feature map re-arrangement strategy commonly implemented in patch classification methodologies (e.g., Boulze et al., 2020; Khaleghian et al., 2021). Our architecture then allows for the generation of predictions for large SAR images in one forward-pass, eliminating the need for patch stitching or mosaicing necessary for patch classification algorithms. This design is also more computationally efficient, as convolutions and other sliding window operations, such as pooling, can be applied over large inputs with comparatively low computational requirements (e.g., Lecun et al., 1998) without negative impact on the produced results. More importantly, our model architecture also shows robustness against Terrain Observation with Progressive Scans SAR (TOPSAR) noise, with very few artifacts that can be attributed to noise. We experiment with different seasonal configurations and present detailed results to investigate how meteorological characteristics can affect the performance of our models under different objectives. We also compare the results of our model with a baseline U-Net model inspired by Stokholm et al.'s (2022) implementation. We use the four-level architecture configuration (standard) presented in their paper and repository, only adapting the very last layer to produce the correct number of classes for our experiments.

## 2. Dataset and Training Targets

The Extreme Earth dataset version 2 is a collection of labels in the form of high-resolution ice charts for 12 Sentinel-1 images in EW mode, that is made available at a 40 m pixel spacing, acquired over the east coast of the Greenland Sea east of Danmarkshavn. The dataset uses Sentinel-1 images of the marginal ice zone to identify a large variety of ice types and was designed and labelled specifically for training automated algorithms. The twelve images are roughly one month apart, one for each month of the year, covering several types of sea ice and different weather conditions throughout 2018. The images were interpreted by expert ice analysts at the Norwegian Meteorological Institute (MET Norway). During the creation of the dataset, polygons were drawn, and sea ice properties assigned primarily according to the interpretation of Sentinel-1 backscatter signatures and supplemented by other remote sensing data that included weather observations, optical images from Sentinel-3 Sea and Land Surface Temperature Radiometer (SLSTR) and Sentinel-2 Multispectral Instrument (MSI). The interpretation was performed at a higher granularity than what is typical for ice charts, resulting in smaller polygons than those used for operational sea ice mapping, as the example in Figure 1 highlights. Additionally, the labels closely correspond with the Sentinel-1 image, unlike typical periodic ice charts, where sea analysts might take into consideration ice drift and potential melting or freezing occurring between the acquisition time of the satellite image and the publication time. Moreover, some ice centres may draw their interpretation assuming that sea conditions will have changed by publication time, adding safety buffers for ice. The Extreme Earth dataset is a more appropriate benchmark than ad-hoc labels or attempts to align published ice charts with SAR images; because it was created by sea ice analysts that ensured that the ice polygons were aligned with the publicly available Sentinel-1 acquisitions.

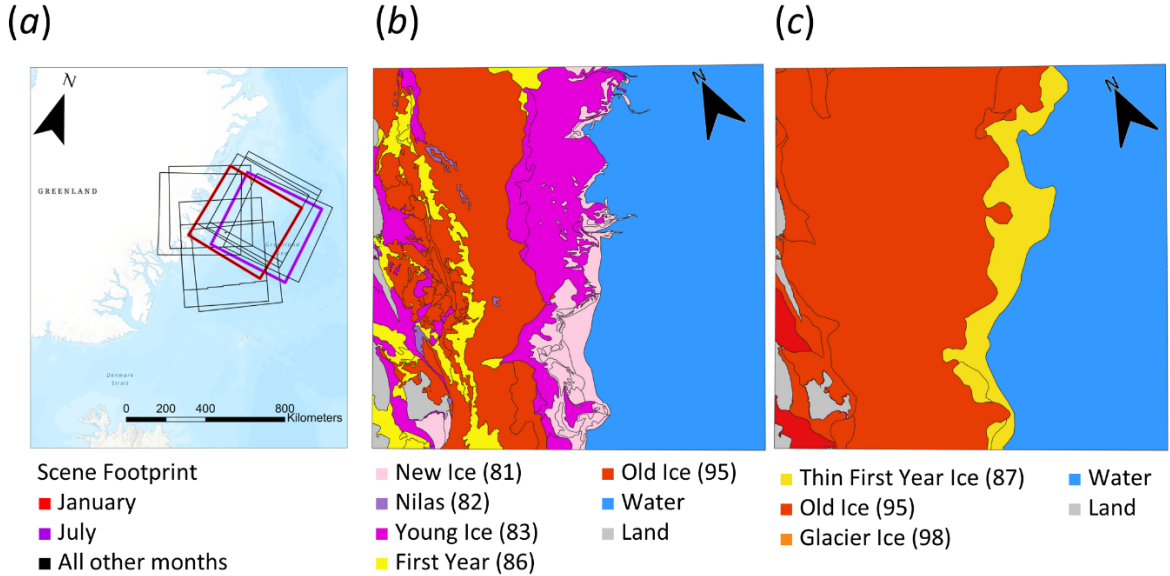

*Figure 1: (a) Extreme Earth labels location. January and December have the same footprint. (b) Extreme Earth ice chart based on a Sentinel-1 image acquired on January 16, 2018; (c) US NIC operational ice chart for January 18, 2018 clipped the Extreme Earth region. Extreme Earth ice charts are drawn with substantially smaller polygons for the purpose of training and testing automated methods. Note the types of ice labelled in (b) are not the same as in (c) given the specific practices at different national ice centres. Digits written in parentheses represent ice type codes in the Egg Code format as made available in Extreme Earth and US NIC operational ice charts.*

In the Extreme Earth dataset, each polygon is assigned a total sea ice concentration, a primary sea ice type, which is the oldest ice in the polygon (coded SA in the Egg Code), a secondary ice type, the second oldest type (coded SB); and partial concentrations for each identified ice type. Polygons are also assigned a form, indicating floe size, and if icebergs are present. We derive an additional label, the dominant ice type, defined as the ice type with the largest partial concentration. We train the models for three targets: a binary classification of sea ice and water; a multiclass classification of oldest ice type; and a multiclass classification of dominant ice type. The dataset is imbalanced, with 47% of the area of all images labelled as open water, while old ice, the next largest class making up only 19% of the total area (see supplementary Tables 2 and 3). For more information about the dataset, as well as its specific location, please refer to Hughes and Amdal (2021).

## 3. Methods

### *3.1. Model architecture and training strategy*

We designed our architecture with the aim of facilitating sea ice characterization by generating outputs for entire Sentinel-1 EW scenes with results that are robust against TOPSAR noise. CNNs are known to be robust against small levels of random noise, however the patch classification of Sentinel-1 EW images is challenging for these models, as a patch might be dominated by systematic banding noise. Therefore, using larger patches for training provides the model with examples of systematic noise overlaid on top of the signal in different neighbouring contexts. After the model is trained, it processes input images with one forward pass, avoiding the generation of potentially inconsistent patch outputs, and the extra post-processing and smoothing necessary to mosaic the patches into larger charts. The architecture we use can be interpreted as having two main parts, an encoder and a decoder (Figure 2). The encoder is a modification of ResNet-18 (He et al., 2016), in which we discarded the last convolutional group. We experimented with keeping the last group and deeper models, including ResNet-50, but we found better results using the adapted ResNet-18 as just described. The decoder is based on the Atrous Spatial Pyramid Pooling (ASPP) module proposed by Chen et al. (2017). The ASPP processes feature maps at multiple scales using atrous (also known as dilated) convolutions, exploiting global and regional spatial context to further boost performance. In atrous convolutions, the field of view of the convolutional kernel increases without requiring extra trainable parameters. The ASPP module provides global context to the final classification layer, which helps with better leveraging spatial context to aid in cases where SAR backscatter is ambiguous over sea ice types. In practice, we modified the ASPP from PyTorch's implementation with an adaptation to allow images with different dimensions during training versus prediction by replacing the global average pooling with average pooling. We configured this average pooling layer using 2x2 kernel size and stride of two. The original global average pooling is a strategy introduced in Chen et al. (2017) to further incorporate global context information in the feature maps used for classification and was inspired by previous methodologies (Grauman and Darrell, 2005; Lazebnik et al., 2006; He et al., 2014). Global context is derived at the dimensions of patches, and because we cannot train the models on full

Sentinel-1 images due to memory limitations, while wanting to preserve the ability to perform prediction on full size SAR images, a global average pooling used in training could negatively affect the model learned parameters. Our modification still incorporates larger spatial context through average pooling and atrous convolutions, albeit without a global pull. Our architecture is based on window-and-stride operations, including (atrous) convolutions, pooling, and upsampling, and therefore, can be used at test time with images of arbitrary size, allowing for the generation of full Sentinel-1 sea ice maps with a single forward pass.

The ResNet-18 encoder has 2.8 M parameters, and is initialized using ImageNet (Russakovsky et al., 2015) weights, i.e., the model was previously trained on ImageNet and is fine-tuned for the sea ice segmentation tasks considered herein. We also experiment with randomly initialized weight in experiments with the dominant ice type. The decoder has 2.8 M parameters and is initialized with random weights. The model has 5.6 M trainable parameters in total. Figure 2 shows the architecture of our model. The encoder downscales the input image by a factor of 16, and the decoder does not change the height or width of the feature maps. Instead, the model resamples the lower resolution decoder output back to the input resolution using bilinear interpolation and then generating outputs with the same spatial dimensions as the input. This is the same strategy used in Chen et al. (2017) as implemented in PyTorch and has the potential to act as a filter avoiding the misclassification of spurious random noise as a different class. Our model architecture and training approach were based on common practices in deep learning studies and our experience and alternative approaches with similar outcomes may exist. Additionally, our algorithm was specifically tailored for analysing Sentinel-1 data using Extreme Earth version 2 labels, and its performance may vary on other missions and ice charts.

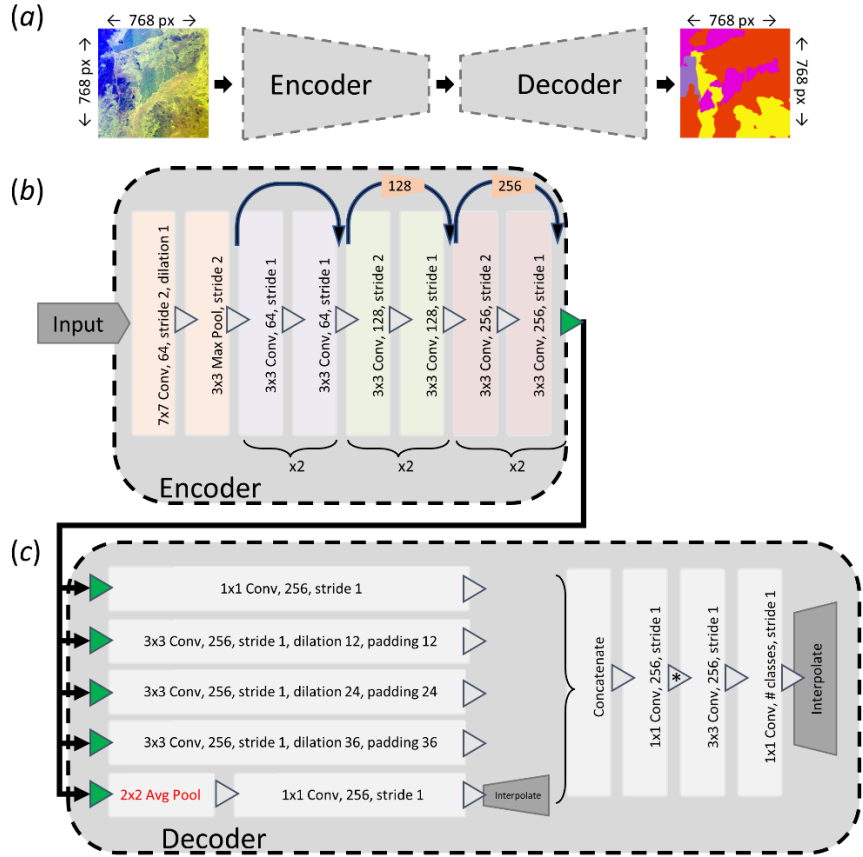

*Figure 2: Model architecture. Max Pool stands for max pooling, Avg Pool for average pooling, Conv for convolution. A) Our model has an encoder and a decoder. The model takes as input arrays with three channels – HH, HV, and incidence angle in all experiments. The output of the model depends on the number of classes based on the specific experiment. The representation in a) also shows the dimensions of the training data for reference (the trained model can be run on entire SAR scenes' dimensions). B) we use ResNet-18 as the encoder for our model. We discarded ResNet's classifier and last convolutional block to reduce the model size and trainable parameters. The orange trapezoids indicate 1x1 Conv with stride 2. C) The decoder of the model is adapted from ASPP. The main modification is highlighted in red (Avg Pool), all the other parameters are kept the same as PyTorch's implementation. Convolution (Conv) layers are followed by batch normalization and activation functions*

### 3.2. Data split and preparation

We conducted several experiments, with varying train/test data splits and classification outputs. Each model used Sentinel-1 Ground Range Detected (GRD) EW swath images as input.

Scenes corresponding with the Extreme Earth ice charts were downloaded from The Copernicus Open Access Hub. We use co-polarized HH and cross-polarized HV channels, and incidence angle as the three input feature channels to the models. Models assign a class (either ice-water or ice type depending on the experiment) for each pixel in a semantic segmentation framework. We processed the Sentinel-1 images using reprojection, thermal noise removal, orbit file correction, radiometric calibration, speckle filtering, and terrain correction using SNAP version 8 following Hughes & Amdal (2021), who observed that terrain correction is crucial for correct geolocation of ascending Sentinel-1 images on the east coast of Greenland. We resampled the Sentinel-1 HH, HV, and incidence angle data from 40 by 40 meters pixel spacing, to 80 by 80 meters pixel spacing, similar dimensions to Stokholm et al. (2022), and converted the raw backscattered values of Sigma0 HH and Sigma0 HV bands to decibels using GDAL (GDAL, 2022).

For each training target, we trained the models on three sets of the data: one training set was composed of images acquired roughly during the freeze-up season (*freeze*), one training set was composed of images in the melt season (*melt*), and one that included all but two images (*all*). Table 1 shows the data split for each set, now denominated experiment group. Images acquired in January and July were not used for training in any experiment and were held out as test images. Having different experiment groups provided extra test scenes for specific groups and allowed us to evaluate the models' generalization for different seasons. For example, October, November, December, February, and March images are not part of the training images for the melt group (i.e., models trained on images acquired during the melt season), and thus, they were used as test images for that group along with the January and July images. Although the meteorological characteristics of melt and freeze seasons are different, ice melt and growth are local processes and there is a transition between seasons. The Sentinel 1 scenes are 400 km wide so there will be variation between freezing close to Greenland's coast and melting at the ice edge. Note, for example, that there is New Ice interpreted on the May scene (part of the melt experiment group), when temperatures in the region varied from -1 ºC to 2 ºC. Having three experiment groups allows us to evaluate if models can achieve better seasonal performance in comparison with a model trained on year-round data. It also allows us to evaluate the performance of models trained with a higher test-to-training data ratio: freeze and melt have seven test and five training scenes. The extra monthly test scenes can indicate if there is consistent deterioration in performance of the models as scenes are farther apart in time from the scenes used in training. Finally, having cross-seasonal evaluation can highlight if different meteorological and physical conditions, as well as different mix of ice types require seasonal machine learning models, or the variety of available training samples acquired throughout all seasons allow a single model to learn the (nonlinear) mapping of SAR backscatter to ice charts during different seasons. Given the small size of this high-resolution benchmark dataset we use a special strategy to generate validation samples. Specifically, we split February, June, August, and December images/charts into their "E" (East) and "W" (West) sections as shown in Figure 3. In doing so, we ensure non-overlapping samples for validation during training (different from test samples completely held out for final evaluation and metric calculation). We split only those four images so that parts of two different scenes are used for validation in *freeze* and *melt* experiments.

Table 1: Images used for training, validation, and testing.

| Month | Train | Validation | Test |
|---|---|---|---|
| January (test) | - | - | all, freeze, melt |
| February-W | all, freeze | - | melt |
| February-E | - | all, freeze | melt |
| March | all, freeze | - | melt |
| April | all, melt | - | freeze |
| May | all, melt | - | freeze |
| June-W | all, melt | - | freeze |
| June-E | - | all, melt | freeze |
| July (test) | - | - | all, freeze, melt |
| August-W | all, melt | - | freeze |
| August-E | - | all, melt | freeze |
| September | all, melt | - | freeze |
| October | all, freeze | - | melt |
| November | all, freeze | - | melt |
| December-W | all, freeze | - | melt |
| December-E | - | all, freeze | melt |

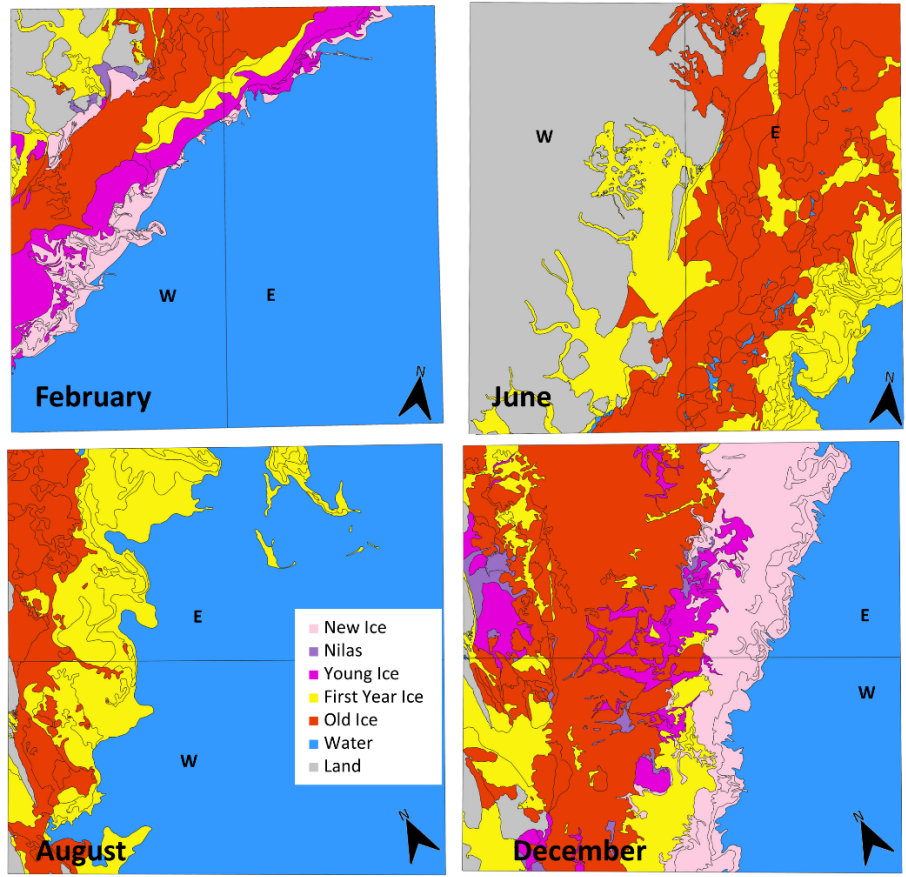

*Figure 3: Oldest ice type (SA) polygons from the Extreme Earth version 2 dataset (Hughes and Amdal, 2021) for selected scenes. We split February, June, August, and December into "W" and "NE" sections as displayed here. W sections were used to generate training samples and E sections were used for validation for some experiments (separate from unseen test images). Refer to Table 1 for more details about experiment groups. The colormap is the same for all images.*

### *3.3. Training and testing*

During training, our data loader randomly extracts patches from the selected Sentinel-1 scenes. We followed the popular strategy to train the models using patches that are smaller than the full Sentinel-1 scenes. The small size of our dataset leading to limited number of samples in conjunction with large memory requirement (for large imagery used during training) made training using full Sentinel-1 scenes impractical. Patches containing more than 30% of invalid pixels were discarded. We considered a pixel invalid if it contained null values as a result of being outside the edges of the SAR image due to image projection, or if it was labelled as land. A patch containing less than 30% invalid pixels was still used for training or validation. In such cases, the invalid pixels in the input data were replaced with channel-dependent mean value of the patch, and the labels were assigned to a category that was ignored during loss and metric computation stages, just like land was ignored. The data loader is adaptive, and the patch sizes can be chosen based on the map units of the raster scenes. We selected square patches of size 61,440 m x 61,440 m, corresponding to 768 x 768 pixels, which was the patch size used by Stokholm et al. (2022) that achieved best performance in their experiments (albeit in a different architecture and for a different target output, i.e., concentration). Taking into consideration the encoder downscaling described previously, the decoder produces outputs of size 48 x 48 pixels that are upsampled back to 768 x 768 using bilinear interpolation to match the original input dimensions. During training, 200 sample patches are extracted from each scene. Therefore, as an example, 2,000 patches were used for training during the experiment group 'all' (200 patches per scene, 10 scenes used). Our data loader normalizes the data (Z-score normalization) before passing it to the models using means and standard deviations calculated from training images listed in Table 1 for each experiment (see supplementary Table 4). The same statistics were used to normalize validation and test data.

For testing and deployment purposes, our models take the full Sentinel-1 scenes as input, enabled by the encoder modification described previously. Sentinel-1 scene dimensions range from roughly 5,000 by 5,000 to 7,000 by 7,000 pixels after projection and resampling. The invalid pixels (null) are replaced with the mean average for each channel, similar to what is done during training. With this strategy, the model generated outputs for each full size Sentinel-1 EW image in less than a minute using an Intel Xeon CPU 3.7Ghz and 128 GB RAM. We repeated each experiment group five times to verify the models' performance metrics against the stochastic nature of training deep networks and show aggregate metrics in the Results section. The only difference in each repetition is that we used different starting seeds for the pseudo random number generators. Different seeds also moved the subimage patch boundaries extracted for training and validation, as well as different starting weights assigned to the model decoder. These differences can affect the final performance on unseen test images as the results in Section 4 show.

We trained the models using cross-entropy loss, and the Adam optimizer (Kingma and Ba, 2014) with a batch size of 32. The learning rate started at $1 \times 10^{-5}$ and was multiplied by 0.1 if the validation loss did not improve for five epochs down to a minimum of $1 \times 10^{-8}$. The models stopped training when the validation loss did not improve for 20 epochs. The models' weights at the lowest validation loss were saved and later used for testing on unseen Sentinel-1 scenes. We trained the models using a NVIDIA RTX A5000 Graphics Processing Unit (GPU). One training epoch took roughly ~0.5 to ~1.2 minutes depending on the number of samples in the experiment group (the *all* group having more samples than *freeze* and *melt*) and model type (ice-water models are marginally faster than stage of development models because of the fewer number of classes). We evaluated the models' performance using traditional metrics, including F1, intersection over union (IoU), and Cohen's kappa. McHugh (2012) provides an extensive discussion around kappa. In this manuscript we highlight different ranges of kappa, including 'moderate' (for kappa values larger than 0.60 and smaller than 0.80), and 'almost perfect' (kappa larger than 0.90). We used PyTorch (Paszke et al., 2019), PyTorch Lightning (Falcon, 2019), Torchmetrics (Detlefsen et al., 2022), and Scikit-Learn (Pedregosa et al., 2011) as the main libraries to develop our models and analysis.

## 4. Results

### *4.1. Ice-water segmentation (binary classification)*

We focus the discussion on test performance (see supplementary Figure 1 for metrics calculated during training). Table 2 shows the test scene-by-scene metric summary for all experiments and all test scenes.

Table 2: Median, minimum, and maximum weighted F1 and Cohen's kappa values for experiment groups computed for all test scenes for ice-water segmentation. Median F1 smaller than 0.95 are highlighted with underline marks. Kappa larger than 0.90 are highlighted in bold. January and July scenes are shared unseen test scenes for all groups.

| Experiment Group | Scene | F1 (weighted) | | | Kappa | | |
|---|---|---|---|---|---|---|---|
| | | Min | Median | Max | Min | Median | Max |
| all | January (test) | 0.98 | 0.98 | 0.98 | 0.95 | **0.95** | 0.96 |
| | July (test) | 0.97 | 0.98 | 0.98 | 0.94 | **0.95** | 0.96 |
| freeze | January (test) | 0.95 | 0.97 | 0.98 | 0.91 | **0.94** | 0.96 |
| | April | 0.98 | 0.98 | 0.98 | 0.96 | **0.96** | 0.96 |
| | May | 0.99 | 0.99 | 0.99 | 0.97 | **0.97** | 0.97 |
| | June-E | 0.97 | 0.97 | 0.98 | 0.80 | 0.82 | 0.86 |
| | June-W | 0.80 | <u>0.82</u> | 0.89 | -0.09 | -0.07 | -0.06 |
| | July (test) | 0.92 | <u>0.94</u> | 0.95 | 0.83 | 0.88 | 0.90 |
| | August-E | 0.85 | <u>0.88</u> | 0.89 | 0.67 | 0.73 | 0.76 |
| | August-W | 0.89 | <u>0.90</u> | 0.92 | 0.67 | 0.71 | 0.76 |
| | September | 0.94 | <u>0.94</u> | 0.94 | 0.74 | 0.76 | 0.76 |
| melt | January (test) | 0.96 | 0.97 | 0.98 | 0.92 | **0.93** | 0.95 |
| | February-E | 0.98 | 0.98 | 0.99 | 0.90 | **0.91** | 0.94 |
| | February-W | 0.97 | 0.98 | 0.98 | 0.93 | **0.96** | 0.96 |
| | March | 0.97 | 0.98 | 0.99 | 0.53 | 0.71 | 0.84 |
| | July (test) | 0.96 | 0.97 | 0.98 | 0.92 | **0.95** | 0.95 |
| | October | 0.96 | 0.98 | 0.98 | 0.92 | **0.94** | 0.96 |
| | November | 0.97 | 0.98 | 0.99 | 0.94 | **0.97** | 0.98 |
| | December-E | 0.99 | 0.99 | 0.99 | 0.96 | **0.97** | 0.98 |
| | December-W | 0.99 | 0.99 | 0.99 | 0.98 | **0.99** | 0.99 |

On average, the models trained in the *all* experiment group performed best on both January and July test scenes with acceptable F1 scores. Therefore, we further analyse these results. Figure 4 shows a model from the *all* group (seed #4) that achieved the highest scoring prediction for January (accuracy, macro, and weighted F1 = 0.98, macro and weighted IoU = 0.96, kappa = 0.96). The red square in Figure 4 highlights a region that is consistently misclassified across the output from most models. This misclassification is attributed to a lower intensity section in the Sentinel-1 image that was labelled as New Ice as shown in Figure 5.

Figure 6 shows the best scoring prediction for July, a model from the experiment group *all* (seed #2) with macro and weighted F1 of 0.98, macro and weighted IoU of 0.97, and kappa of 0.96. For July, there is a thin line of error along the boundary between water and ice, which can potentially be attributed to segmentation on the lower resolution of intermediate feature maps generated by the encoder as described in Section 3.1. However, as discussed above, training the segmentation on lower resolution feature maps makes the model less sensitive to TOPSAR noise. Figure 7 shows the confusion matrices for results shown in Figures 4 and 6. The confusion matrix does not indicate any significant model bias. The median weighted F1 scores obtained on the January test scene for the three experiment groups are equal to or larger than above 0.97. The *freeze* experiment group is the only one with a median below 0.95 for July. This shows strong potential for automation of ice-water segmentation using our method.

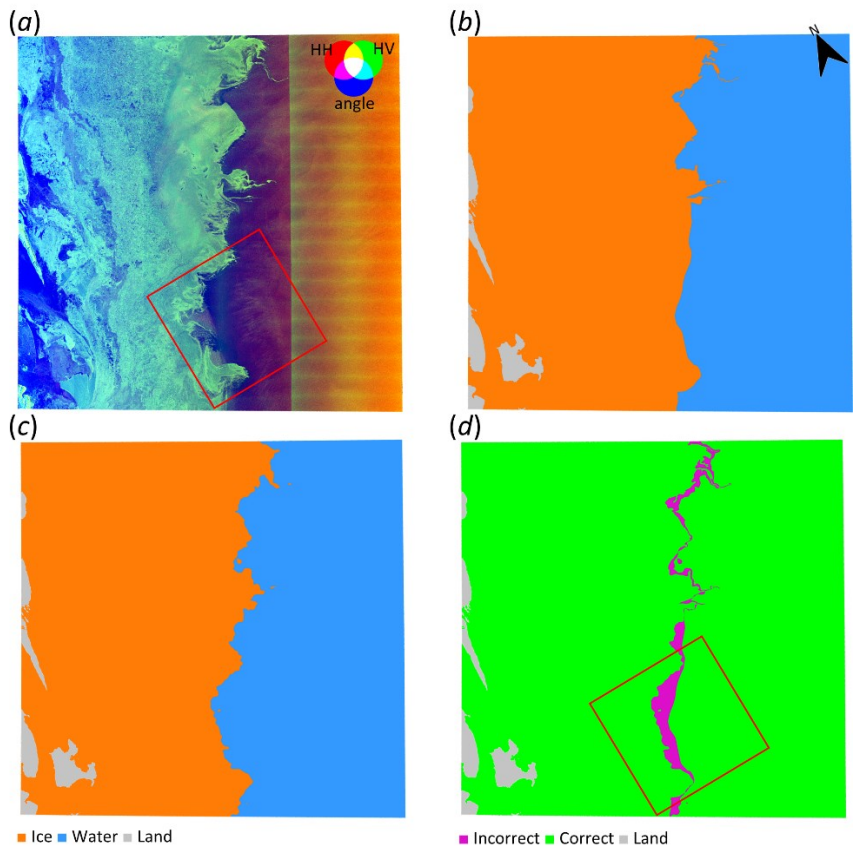

*Figure 4: Scene prediction example for January test image (held out during training). This figure shows the best scoring model for the January scene (experiment group all, seed #4). (a) Model input, (b) rasterized labels from the Extreme Earth dataset, (c) model output, and (d)*

*prediction error. The land labels are ignored during metric computation. The red square highlights a location on which our models consistently classify New Ice as water (Figure 5).*

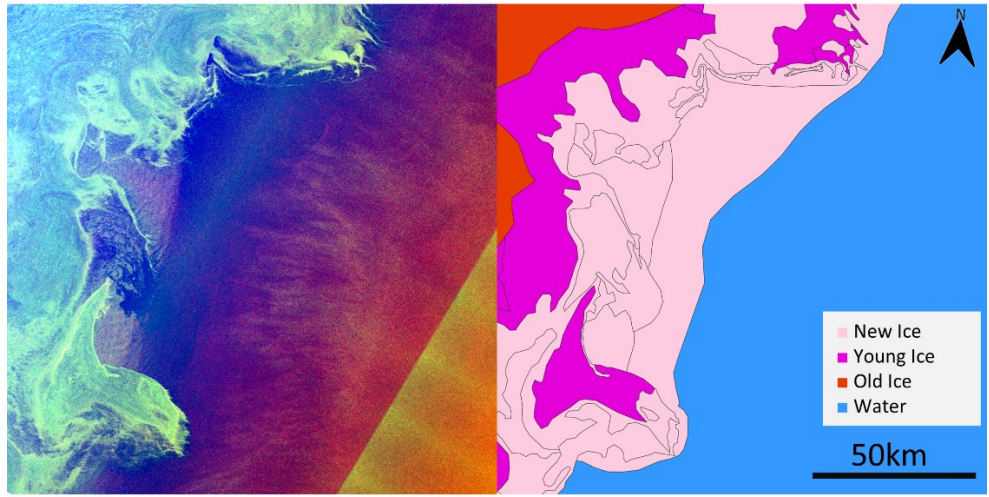

*Figure 5: A section of Sentinel-1 image and Extreme Earth ice chart polygons for January. The image on the left uses the same colour scheme as Figure 4. Dark New Ice with low backscatter intensity coinciding with high sub-swath edge noise is misclassified as water by our models. The location of this inset is shown in Figure 4.*

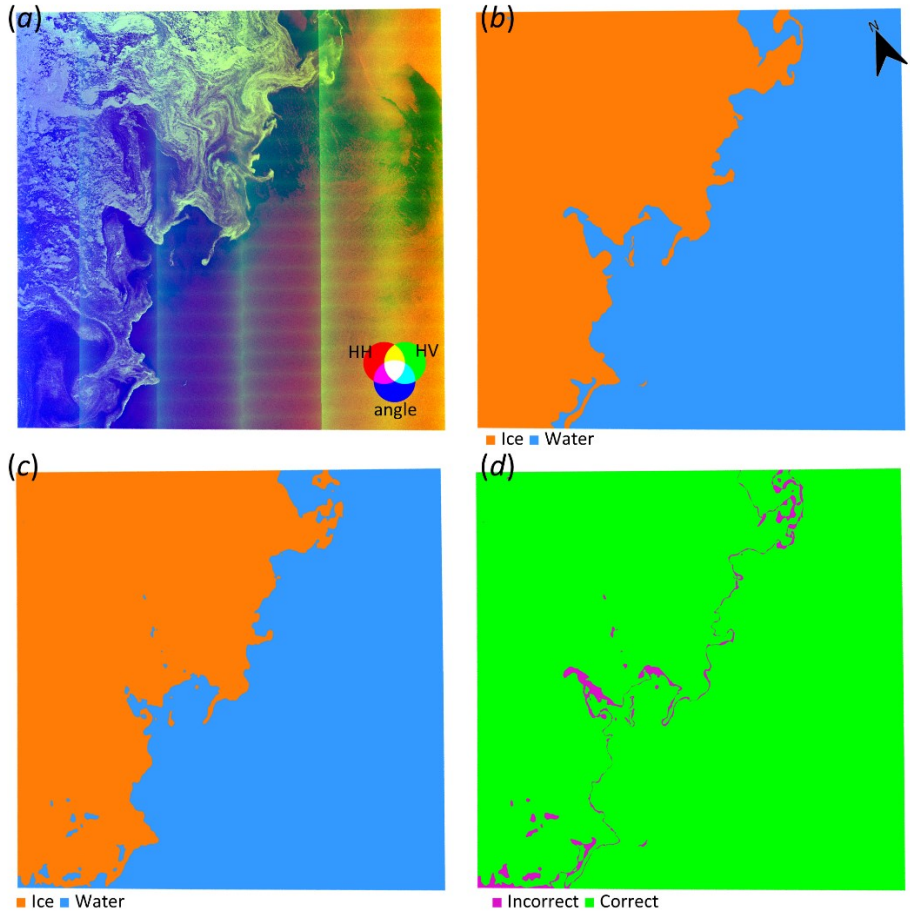

*Figure 6: Scene prediction example for July. This figure shows the best scoring model for the July scene (experiment group all, seed #2). (a) Model input that shows strong banding noise, (b) rasterized labels from Extreme Earth, (c) model output, and (d) comparison between labels and model output.*

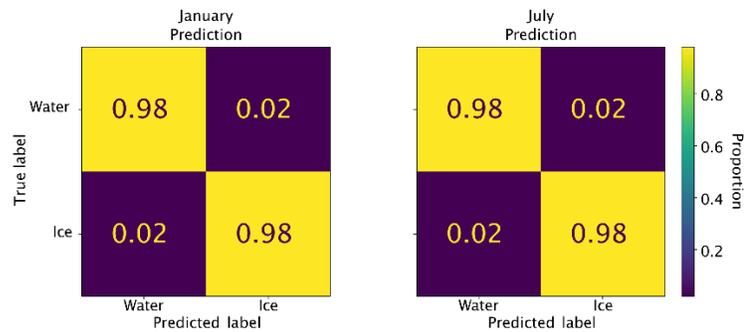

*Figure 7: Confusion matrix for the results shown in Figure 4 and 6. Values are normalized per row (true label) and the colormap is the same for both figures.*

Results in Table 2 show only two test scenes with weighted F1 below 0.90: June-W and August-E, both for the *freeze* experiment group. This is not surprising, as those models have been trained only on images acquired during the freeze up season and underperform on images during and after the melt period (that cover surface and snow melt). Although the poor performance for June-W can partially be attributed to a region of smooth First Year Ice being classified as water by the model, *freeze* models also show weaker performance in August. Only five out of 20 tests in Table 2, all from the *freeze* experiment group testing on *melt* months, have maximum weighted F1 below 0.98. The *all* experiment group weighted F1 are all above 0.97. This is generally acceptable and shows strong potential for automating ice-water segmentation using our method if trained on images acquired year-round.

### *4.2. Stage of Development (Type) Classification: Oldest Type (SA)*

Table 3 shows a scene-by-scene test metric summary for all experiments and all test scenes (see supplementary Figure 2 for metrics calculated during training). The median of weighted F1 is equal to or larger than 0.6 for January and July, the main test scenes. This lower performance on test scenes is an indication of overfitting, which may indicate some subjectivity of the ice types assigned in ice charts by the expert analyst. Another potential cause might be the small number of training samples compared to the complexity of type classification.

Table 3: Median, minimum, and maximum weighted F1 and Cohen's kappa values for experiment groups computed for all test scenes for oldest ice type segmentation. F1 Medians higher than 0.80 are highlighted in bold. Kappa higher than 0.6 are highlighted in bold.

| Experiment Group | Scene | F1 (weighted) | | | Kappa | | |
|---|---|---|---|---|---|---|---|
| | | Min | Median | Max | Min | Median | Max |
| all | January (test) | 0.61 | 0.62 | 0.66 | 0.49 | 0.50 | 0.54 |
| | **July (test)** | 0.79 | **0.82** | 0.82 | 0.66 | **0.67** | 0.70 |
| freeze | January (test) | 0.59 | 0.64 | 0.67 | 0.48 | 0.53 | 0.56 |
| | April | 0.45 | 0.49 | 0.52 | 0.38 | 0.40 | 0.42 |
| | **May** | 0.85 | **0.86** | 0.88 | 0.57 | **0.60** | 0.63 |
| | June-E | 0.17 | 0.19 | 0.22 | 0.03 | 0.04 | 0.08 |
| | June-W | 0.44 | 0.49 | 0.55 | -0.12 | -0.09 | -0.02 |
| | July (test) | 0.58 | 0.6 | 0.68 | 0.33 | 0.36 | 0.49 |
| | August-E | 0.58 | 0.63 | 0.67 | 0.23 | 0.33 | 0.42 |
| | August-W | 0.75 | 0.77 | 0.79 | 0.34 | 0.38 | 0.46 |
| | **September** | 0.86 | **0.86** | 0.87 | 0.46 | 0.47 | 0.49 |
| melt | January (test) | 0.62 | 0.66 | 0.71 | 0.46 | 0.55 | 0.62 |
| | **February-E** | 0.91 | **0.92** | 0.92 | 0.62 | **0.67** | 0.68 |
| | February-W | 0.62 | 0.64 | 0.65 | 0.53 | 0.57 | 0.59 |
| | March | 0.55 | 0.57 | 0.61 | 0.34 | 0.36 | 0.41 |
| | **July (test)** | 0.82 | **0.84** | 0.86 | 0.67 | **0.72** | 0.76 |
| | October | 0.65 | 0.74 | 0.78 | 0.33 | 0.47 | 0.54 |
| | November | 0.72 | 0.73 | 0.76 | 0.62 | **0.63** | 0.66 |
| | December-E | 0.53 | 0.54 | 0.55 | 0.41 | 0.45 | 0.46 |
| | December-W | 0.65 | 0.67 | 0.68 | 0.58 | 0.59 | 0.60 |

Much like the results observed during the binary experiments, models trained in the *freeze* experiment group exhibit weaker performance when used to predict scenes in the *melt* season. *Melt* models are also weaker when evaluated against the *freeze* scenes, but the gap is smaller. Curiously, although the three experiment groups have similar performance for the January test scene, the best performance is obtained by a model trained on the *melt* experiment group (seed #3).

The *melt* experiment group had the best performing models for the January test scene (seed #3, accuracy of 0.73, macro F1 of 0.38, weighted F1 of 0.71, macro IoU of 0.32, weighted IoU of 0.64, and kappa of 0.62). A model from the *melt* group was also the best performer for the July test scene (seed #1, accuracy of 0.85, macro F1 of 0.47, weighted F1 of 0.86, macro IoU of 0.41, weighted IoU of 0.78, and kappa of 0.76). Figure 8 shows the results obtained for January using *melt* seed #3 and July using *melt* seed #1. Results for January in Figure 8 and Figure 9 show that the model misclassified most of the New Ice and Nilas labelled in the Extreme Earth dataset, both having lower support compared to other classes. As a result, both *melt* models tend to not predict New Ice, Nilas, and Young Ice, which is expected as the number of samples for those classes are very small for the *melt* training months. The model trained on the *freeze* experiment group with the best scores for January based on weighted F1 scores (seed #3, accuracy 0.67, macro F1 of 0.43, weighted F1 of 0.67, macro IoU of 0.34, weighted IoU of 0.58, and kappa of 0.54) shows weaker performance for New Ice, with F1 score of 0.46, also not predicting Nilas well.

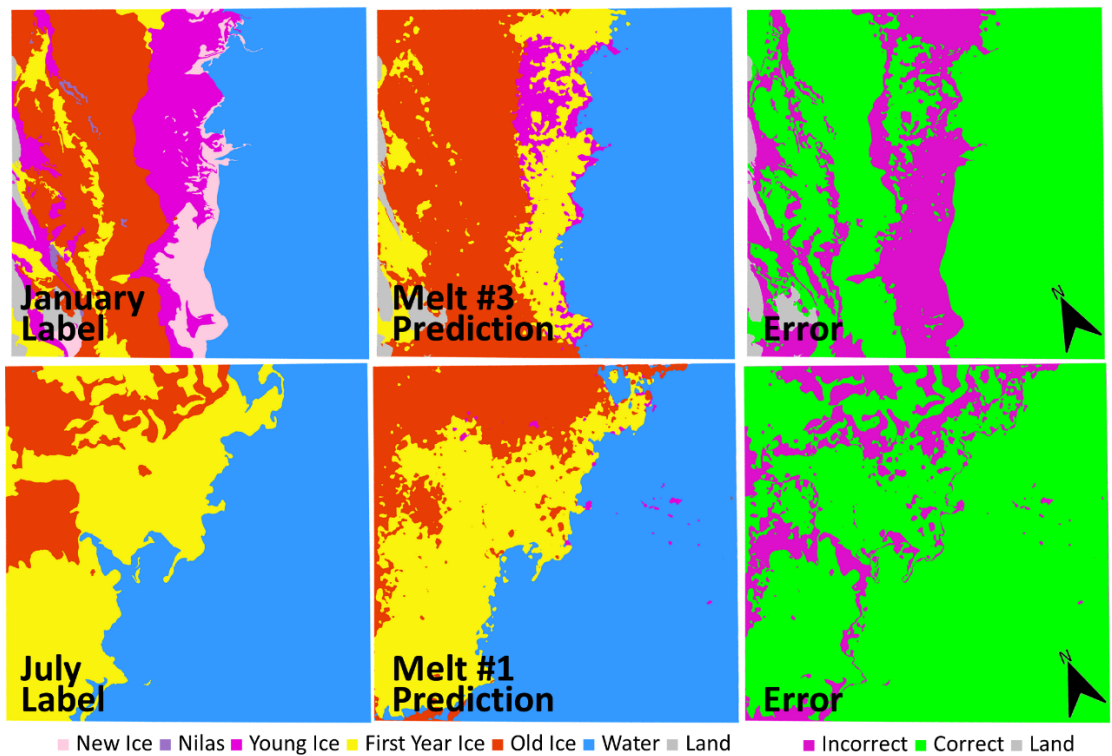

*Figure 8: Output examples for the best oldest ice type predicting model for the main test scenes. Top row shows label, prediction, and error for the January scene, with results obtained by melt #3. Bottom row shows the corresponding July images, with results obtained by melt #1.*

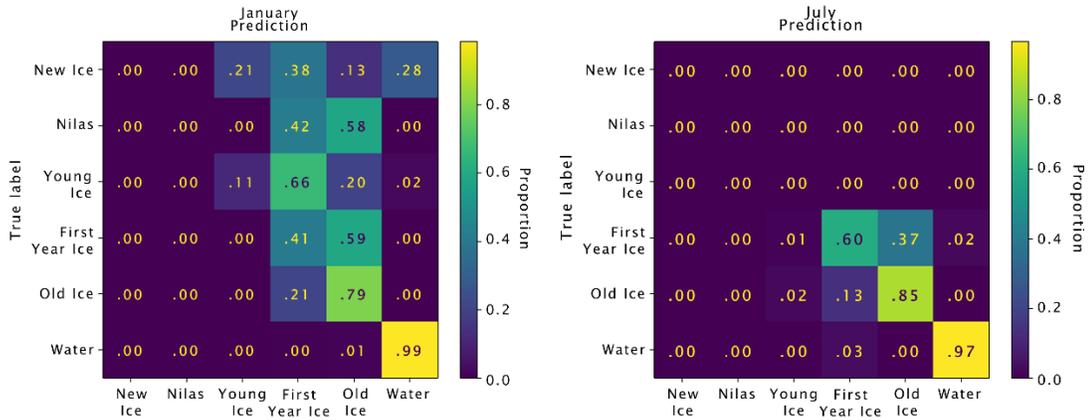

*Figure 9: Confusion matrix for the results shown in Figure 8. January results show that the model tends to not predict New Ice and Nilas. Some young ice, which is typically darker and smoother than ice that has grown thicker, is mistaken for water in January. Absence of New Ice, Nilas, and Young Ice in July is advantageous for the model's metric computation. Values are normalized per row (true label).*

Despite relative success for binary segmentation of ice/water on S1 EW images, determining oldest type (as interpreted by an expert analyst) remains challenging due to the models, potentially due to subjectivity of oldest type interpretation, ambiguity of the SAR backscatter, as well as the relatively small training dataset compared to the complexity of the task and number of classes. The F1 score of the models for water prediction remains high.

### *4.3. Stage of Development (Type): Dominant Ice Type*

Finally, we repeat the same experiments for the *dominant* ice type. The rationale for a separate set of experiments for this group is that the oldest type in each polygon is not necessarily the ice type with highest concentration, and therefore, may lead to model confusion during training. The downside of looking at the dominant type, however, is that some of the oldest types present in polygons will be ignored during training, as it is unclear which pixel refers to what kind of ice type.

Table 4 shows a scene-by-scene test metric summary for all dominant ice type experiments. The performance of the models is generally improved in comparison with the oldest ice type experiments. Figure 10 shows the best performing model for January (*freeze*, seed #4, accuracy of 0.83, macro F1 of 0.60, weighted F1 of 0.83, macro IoU of 0.50, weighted IoU of 0.74, and kappa of 0.76) and for July (all, seed #1, accuracy of 0.77, macro F1 of 0.33, weighted F1 of 0.81, macro IoU of 0.29, weighted IoU of 0.73, and kappa of 0.61).

Table 4: Median, minimum, and maximum weighted F1 and Cohen's kappa values for experiment groups computed for all test scenes for dominant ice type segmentation. F1 medians higher than 0.80 are highlighted in bold. Kappa medians higher than 0.60 are highlighted in bold.

| Experiment Group | Scene | F1 (weighted) | | | Kappa | | |
|---|---|---|---|---|---|---|---|
| | | Min | Median | Max | Min | Median | Max |
| all | January (test) | 0.70 | 0.77 | 0.79 | 0.60 | **0.69** | 0.71 |
| | July (test) | 0.71 | 0.77 | 0.81 | 0.47 | 0.53 | 0.61 |
| freeze | **January (test)** | 0.78 | **0.82** | 0.83 | 0.70 | **0.74** | 0.76 |
| | April | 0.57 | 0.63 | 0.65 | 0.40 | 0.46 | 0.53 |
| | **May** | 0.91 | **0.92** | 0.93 | 0.66 | **0.71** | 0.76 |
| | June-E | 0.61 | 0.69 | 0.71 | 0.19 | 0.27 | 0.29 |
| | June-W | 0.42 | 0.59 | 0.74 | -0.07 | -0.04 | -0.01 |
| | July (test) | 0.54 | 0.58 | 0.63 | 0.31 | 0.35 | 0.39 |
| | August-E | 0.59 | 0.69 | 0.71 | 0.27 | 0.40 | 0.42 |
| | August-W | 0.77 | 0.79 | 0.79 | 0.36 | 0.42 | 0.42 |
| | **September** | 0.86 | **0.88** | 0.88 | 0.45 | 0.50 | 0.52 |
| melt | January (test) | 0.57 | 0.61 | 0.61 | 0.47 | 0.51 | 0.52 |
| | **February-E** | 0.90 | **0.91** | 0.94 | 0.60 | **0.64** | 0.77 |
| | February-W | 0.59 | 0.62 | 0.64 | 0.49 | 0.54 | 0.54 |
| | March | 0.53 | 0.55 | 0.57 | 0.24 | 0.29 | 0.30 |
| | July (test) | 0.68 | 0.77 | 0.81 | 0.44 | 0.53 | 0.60 |
| | October | 0.73 | 0.78 | 0.78 | 0.49 | 0.58 | 0.59 |
| | November | 0.71 | 0.74 | 0.77 | 0.61 | **0.65** | 0.71 |
| | December-E | 0.48 | 0.50 | 0.53 | 0.38 | 0.42 | 0.46 |
| | December-W | 0.60 | 0.63 | 0.67 | 0.52 | 0.55 | 0.60 |

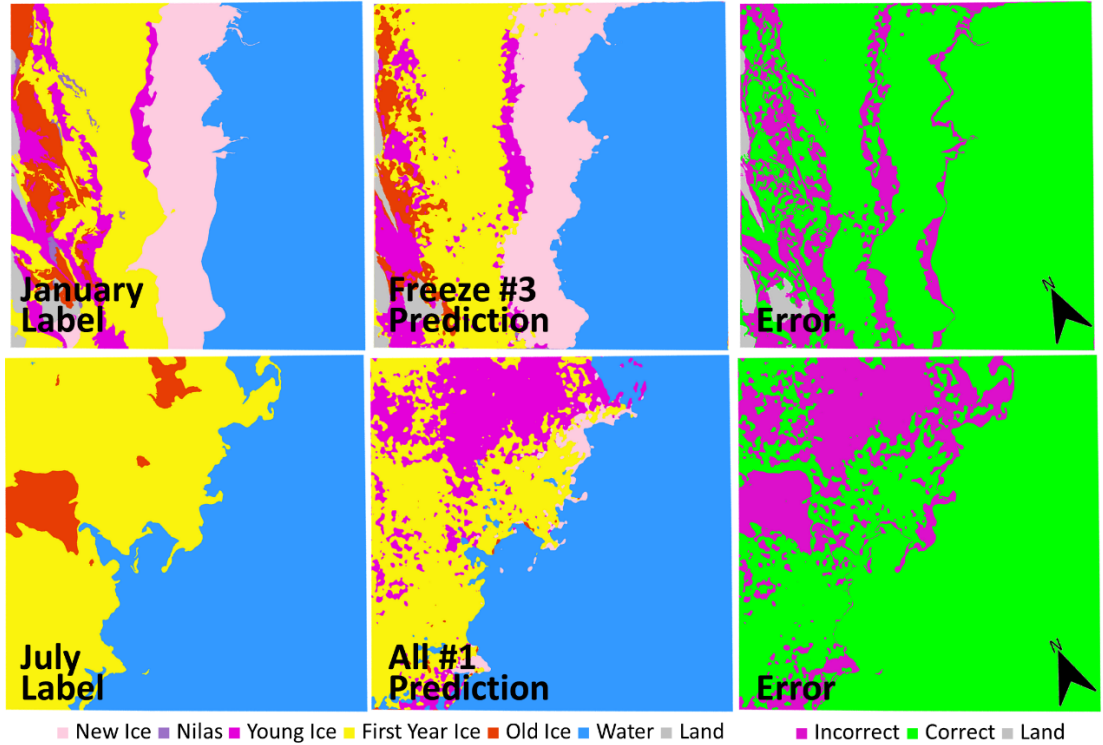

*Figure 10: Output examples for the best dominant ice type predicting model for the main test scenes. Top row shows label, prediction, and error for the January scene, with results obtained by freeze #3. Bottom row shows the corresponding July images, with results obtained by all #1. The same challenging New Ice region highlighted in Figure 5 is misclassified again in January.*

The *freeze* #3 model tends to not predict Nilas in January. The support for Nilas has decreased significantly, as the largest polygon containing Nilas in the January image is now labelled as New Ice, since that is the "dominant" ice type in the polygon. Figure 11 shows that the best performing model identified 85% of New Ice correctly with the dominant ice type set up in January. However, in July, the *all* #1 model classifies most of the Old Ice (71%) as First Year Ice. These numbers further highlight the complexity of automating ice type classification, given that ice charts of types are generated with uncertainty, resulting in polygons of mixed types.

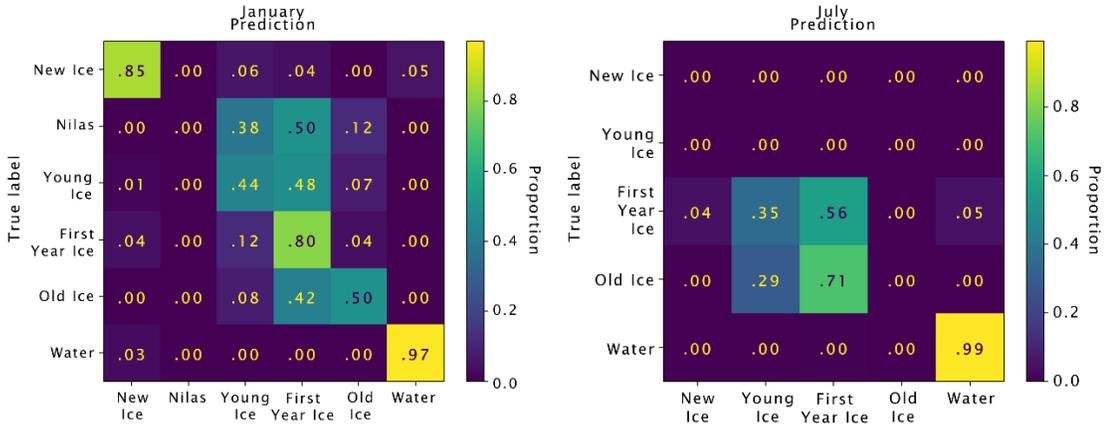

*Figure 11: Confusion matrix for the results shown in Figure 10. Values are normalized per row (true label).*

### *4.4. Comparison of Random Initialization and Pre-Trained models for Dominant Ice Type*

We also used the dominant ice type to evaluate if pre-trained models provide significant advantage in comparison with models initialized with random weights. The advantage of pre-trained models is evident during training, with better scoring metrics for both train and validation sets obtained in fewer epochs (see supplementary Figure 3). This advantage is not so clear once the models are trained and used for full scene prediction on unseen images. Table 5 shows comparison metrics for the test scenes. Pre-trained models achieve higher weighted F1 median performance for the January test scene, but randomly initialized models result in higher median values for July.

Table 5: Minimum, median, and maximum weighted F1 values for test scenes for dominant ice type model initialization experiments.

|                    |                  | F1 (weighted) | | |
|--------------------|------------------|------|--------|------|
| Experiment Group   | Scene            | Min  | Median | Max  |
| all: pre-trained   | January (test)   | 0.70 | 0.77   | 0.79 |
|                    | July (test)      | 0.71 | 0.77   | 0.81 |
| all: random init   | January (test)   | 0.66 | 0.75   | 0.76 |
|                    | July (test)      | 0.75 | 0.82   | 0.85 |

Table 6 and Figures 12 and 13 show a comparison between the model with highest scores for January and July. Figure 12 shows similarities between the predictions from the pre-trained model and the predictions from the model initialized with random weights. Noticeably, the band for Young Ice is narrower for the pre-trained model, which has better correspondence with the Extreme Earth ice charts. Results for the randomly initialized weights seem to generate more homogenous, larger regions for January and July. This apparent reduction in small region predictions is also visible in Figure 13 (July). However, the randomly initialized model shows a linear streak of First Year Ice inside Young Ice (left side) that we interpret is caused due to Sentinel-1's sub-swath noise. Furthermore, the randomly initialized model also exhibits a somewhat weaker skill in separating water from ice, which is also reflected in the metrics. The precision, recall, and F1 score for water for the pre-trained model is, respectively, 0.97, 0.99, and 0.98. The model initialized with random weights achieves 0.94, 0.97, and 0.95 for the same metrics.

Table 6: Scene metrics for Figures 15 and 16.

| Scene | Training mode | Seed # | Accuracy | Macro F1 | Weighted F1 | Macro IoU | Weighted IoU |
|---|---|---|---|---|---|---|---|
| January | pre-trained | 1 | 0.79 | 0.54 | 0.79 | 0.44 | 0.71 |
| January | random initialization | 3 | 0.75 | 0.51 | 0.76 | 0.41 | 0.67 |
| July | pre-trained | 1 | 0.77 | 0.33 | 0.81 | 0.29 | 0.73 |
| July | random initialization | 3 | 0.83 | 0.38 | 0.85 | 0.33 | 0.77 |

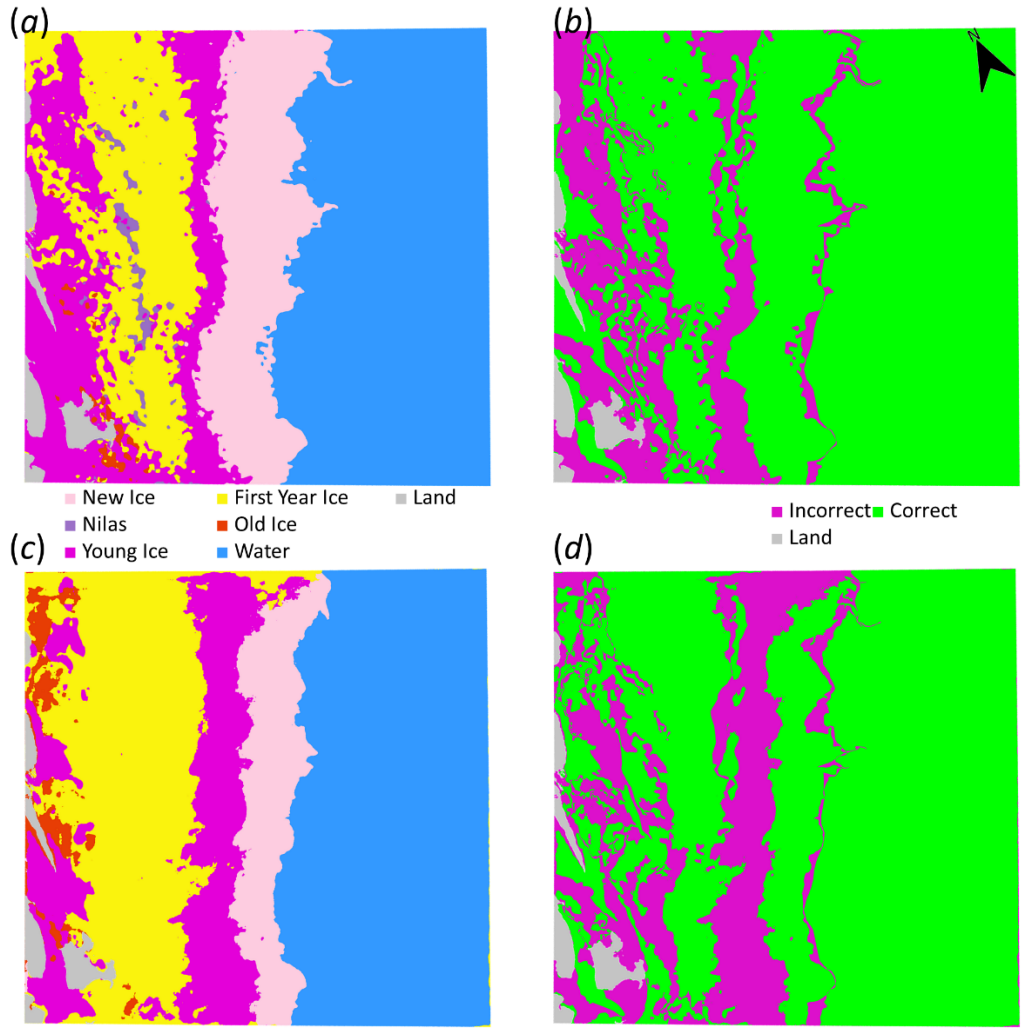

*Figure 12: Comparison of the models with highest metrics scores for January with different training modes. (a) Prediction from model initialized using pre-trained weights and its error in (b). (c) Prediction for model initialized with random weights and its error in (d). (a) and (c) use the same colormap, just like (b) and (d). The comparison and errors are calculated based on Extreme Earth labels.*

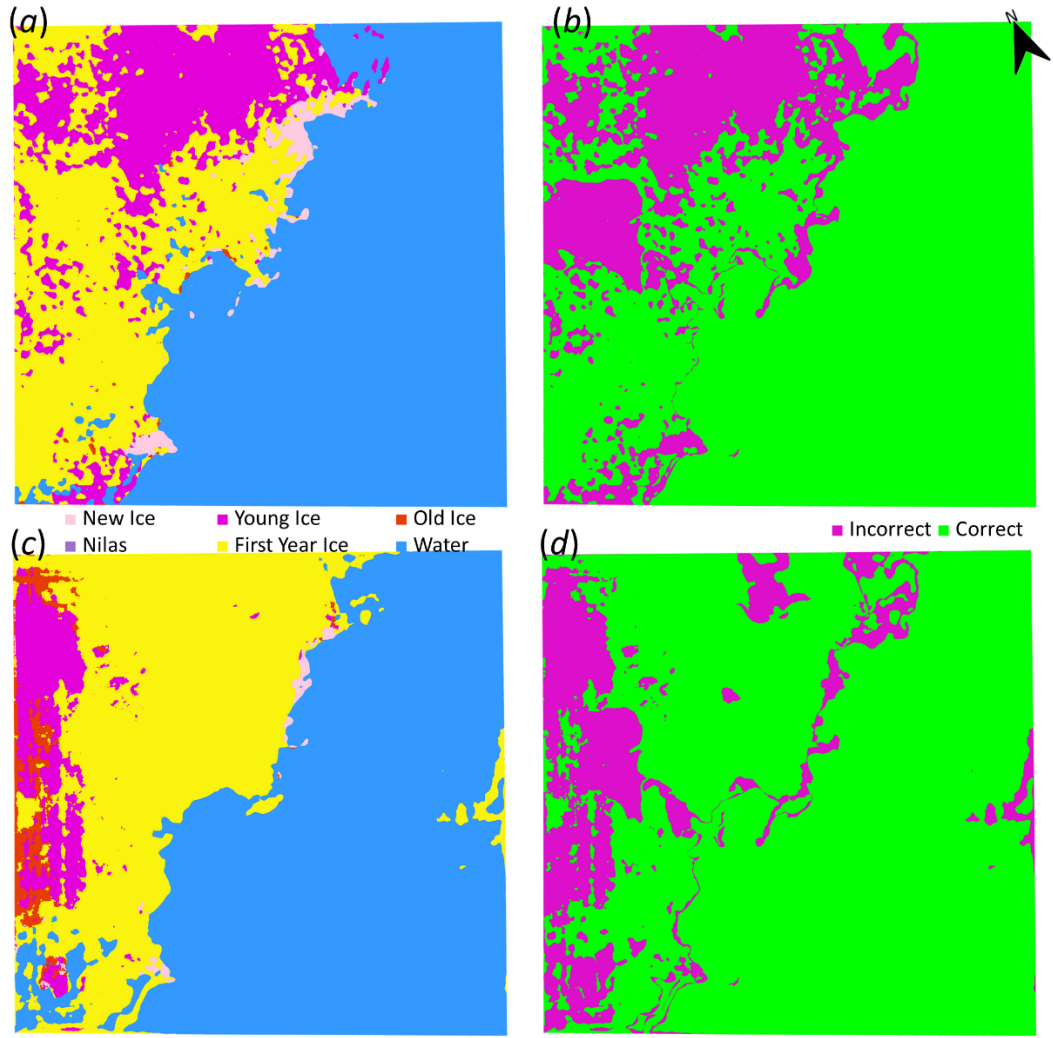

*Figure 13: Comparison of the models with highest metrics scores for July with different training modes. (a) Prediction from model initialized using pre-trained weights and its error in (b). (c) Prediction for model initialized with random weights and its error in (d). (a) and (c) use the same colormap, just like (b) and (d). The comparison and errors are calculated based on Extreme Earth labels.*

### 4.5. Comparison with U-Net

To our knowledge, no other sea ice segmentation paper in the literature reported performance on the Extreme Earth Version 2 dataset. Therefore, we cannot directly compare our results to other studies, which frequently use privately owned imagery and labels for training. Furthermore, the number of ice classes are different in most studies (e.g., in our case, six SoD classes), and the polygons are drawn with different objectives and at different sizes, which make comparison difficult. Therefore, we compare the performance of our model against a baseline U-Net, arguably the most popular semantic segmentation model architecture. As mentioned in the Introduction, we implement the baseline U-Net model by replicating the architecture described in Stokholm et al.'s (2022), which was focused on the classification of Sea Ice Concentration. We used the same hyperparameters described in Section 3.3, except for the batch size. We had to reduce the batch size to 16, as U-Net used more memory during training. We also repeated the experiments five times, considering only the *all* experiment group for this comparison. Table 7 shows a summary comparison between our model and U-Net. For easier reference, results in Table 7 repeat metrics for ice-water (Table 2) and oldest ice type (Table 3). Our model achieves higher weighted F1 for all conditions for the ice-water experiment. For the oldest ice type, our model obtained higher metrics for all July categories, and tied in the median value for January, with the baseline U-Net achieving higher metrics for the maximum weighted F1 by 0.02 points, even though the median scores are tied.

Table 7: Minimum, median, and maximum weighted F1 values for test scenes comparing our model with U-Net. Best results for each target type and scene are bolded.

| Target | Model | Scene | F1 (weighted) | | |
| --- | --- | --- | --- | --- | --- |
| | | | Min | Median | Max |
| Ice-water | Ours | January | **0.98** | **0.98** | **0.98** |
| | | July | **0.97** | **0.98** | **0.98** |
| | U-net | January | 0.97 | 0.97 | 0.97 |
| | | July | 0.92 | 0.93 | 0.94 |
| Oldest Ice Type | Ours | January | **0.61** | 0.62 | 0.66 |
| | | July | **0.79** | **0.82** | **0.82** |
| | U-net | January | 0.51 | 0.62 | **0.68** |
| | | July | 0.76 | 0.78 | 0.80 |

Figure 14 shows the error map for January for the best performing U-Net. In general, we observe U-Net also producing results generally robust against Sentinel-1 noise. However, although hard to observe at this print scale, there are some misclassifications that might be attributed to noise. The model also seems to have a slightly larger band of border artifact misclassifications.

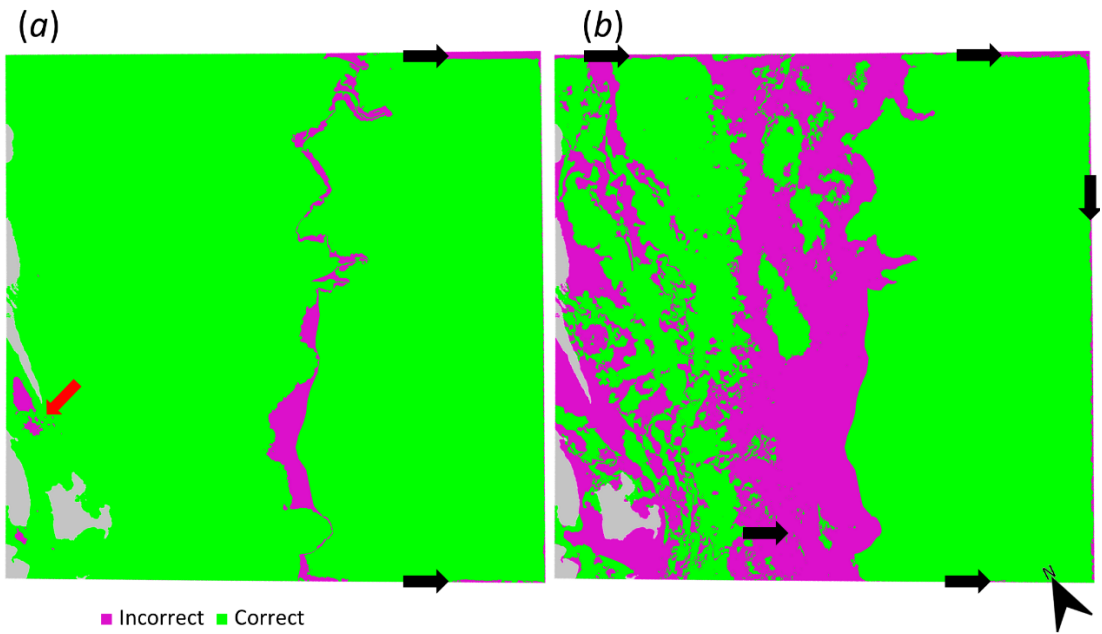

*Figure 14: Error maps for the best performing baseline U-Net models in January. (a) Ice-water error map. (b) Oldest ice type error map. The red arrow points to a challenging dark ice region incorrectly classified by U-Net that is correctly classified by our models. The black arrows point to misclassifications by the U-net likely caused by noise and border effects, that are mostly classified correctly by our models.*

Although our proposed model architecture contains more parameters (5.6 m trainable parameters) than the baseline U-Net (165 k trainable parameters), in general it can process full Sentinel-1 scenes faster. Aggregate values across 16 Sentinel-1 scenes (the 12 scenes discussed in this manuscript considering Feb, Jun, Aug, and Dec are used twice) show that U-Net processes the data in 47.4 ± 30 seconds, while our model architecture processes the data in 11.4 ± 3 seconds. These indicate a time decrease of roughly 76%. The main reason for the speedup is that our model architecture performs several operations on lower resolution feature maps, as discussed in Section 3.1, while U-Net processes the data using larger feature maps. The trade-off between necessary spatial resolution and available time might be an important factor to be considered when choosing different architectures.

## 5. Discussion and Future Work

### *5.1. Ice-Water and Stage of Development (Type)*

The results show that our models have high potential to automate sea ice-water segmentation, with F1 and IoU metrics consistently higher than 0.97 for test scenes, even when using few images for training. We identified one region that is challenging for our models in Figure 6. The "dark" New Ice is barely discernible. Ice experts in this case, used temperature, wind, and observation of solid area of wave dampening in the SAR image to perform their interpretation, in addition to images of the previous day, when water was less calm and easier to discriminate against New Ice. We also observed that the limited number of training samples made it likely to overfit deeper models. We obtained better results for test scenes with our modified, shallower, ResNet-18 excluding the fourth convolutional block, than with the larger ResNet-50 during initial experimentation. While we provide a caveat that the results of different models trained on different datasets are not directly comparable, we contextualize our results against a recent study here. Our results using an open benchmark dataset and Sentinel-1 EW images (which contain systematic banding noise in the cross-polarized channel) provide higher accuracy than those presented by Chen et al. (2023) using RADARSAT-2 images. For January, for example, the average accuracy is 98.0% for the *all* experiment group. If we include July, the average accuracy drops to 97.9% (averaged two scenes and five experiments). Chen et al.'s average accuracy across different model combinations varies from 90.66% to 91.41%. Although their test set contains 14 RADARSAT-2 scenes, Chen et al.'s training set with 21 scenes is at least two times larger than ours. Khaleghian et al. (2021) reported Overestimation, Underestimation, and Integrated Ice Edge Error (IIEE) (Goessling et al., 2016) for the same January scene labelled in the Extreme Earth Version 2. We calculated the mean values and standard deviation for Overestimation, Underestimation, and IIEE with the five results in the *all* experiment group. We obtained smaller values for Overestimation (with mean 1475 and standard deviation ± 374 km$^2$ vs 3766 km$^2$ from Khaleghian et al. (2021)), Underestimation (2444 ± 324 km$^2$ vs 4646 km$^2$), and IIEE (3919 ± 164 km$^2$ vs 8412 km$^2$). Even though the January scene includes the section of New Ice that is challenging for our models as discussed above, our performance metrics are better compared to those reported by Khaleghian et al. (2021). Although we are comparing these metrics, we note that they were calculated based on our geo-referencing and processing, and we included the values reported by Khaleghian et al. (2021). Projection or other distortions might cause a difference in results. Khaleghian et al. (2021) dataset consisted of 31 Sentinel-1 images. We note again, however, comparison with other studies serves a qualitative purpose rather than direct model architecture comparison, as differences in training and input data, label resolution, number of available samples, and labelling practices makes the values of a direct quantitative comparison limited.

    Although we obtained metrics close to 1.0 for ice-water discriminations, the multiclass classification task for both the oldest ice type and dominant ice type did not perform as well. Weighted F1 for the January test scene for classifying the oldest ice type ranged from 0.59 (worst performing model, *freeze* #5) to 0.71 (best performing model, *melt* #3). Analogous values for July are 0.58 (*freeze* #3) and 0.86 (*melt* #1). Performance metrics for the same test scenes but classifying for dominant ice type, ranged from 0.57 (*melt* #4) to 0.83 (*freeze* #4) for January and 0.54 (*freeze* #5) to 0.81 (*all* #1 and *melt* #1). The performance metrics of the models on the test set generally fall below 0.8 when the outputs of the models are compared for SoD segmentation. Results also show potential when adopting a slightly different strategy to label polygons based on the dominant SoD rather than the oldest ice. Considering only the *all* experiment group, there is an increase in median weighted F1 for January from 0.62 to 0.77. This increase in median weighted F1 is not observed for July, with results showing higher scores for the oldest ice type (0.82) than dominant ice type (0.77). However, the average of the median weighted F1 for all test scenes (average of 'median' columns in Tables 6 and 7) increased from 0.66 (oldest ice type) to 0.71 (dominant ice type). One of the reasons for the improvement in performance is that the ice type with higher concentration, i.e., dominant ice type, is what represents most of the ice in a single polygon. Therefore, the features maps extracted for each polygon will have a more direct relationship with the polygon label. In our understanding, training on a hypothetical benchmark dataset labelled at a pixel-by-pixel level, with each pixel representing only the sea ice type present in that pixel, has the potential to improve the quality of the trained models, although we understand producing such a dataset poses different challenges. Moreover, there is usually not a direct discriminatory relationship between radar backscatter and all sea ice types, especially when correlating sea ice type and a single frequency C-band. Casey et al. (2016) observed that

during advanced melt, the contrast between first-year ice and multi-year ice is generally small at C- and L-band.

Results also show that our model generally fails to classify Nilas as labelled in Extreme Earth. This is likely due to the very small number of Nilas samples in the dataset, as well as a weak physical separability from other sea ice types or water as due to radar backscatter properties as discussed in the paragraph above. Nilas does not seem to have been considered a separate sea ice class in several other machine learning studies (e.g., Boulze et al., 2020; Khaleghian et al., 2021; Singha et al., 2021). Increasing the number of samples, as well as improving the class balance has the potential to increase the accuracy of models.

### *5.2. Seasons and number of samples*

An analysis of the results obtained by different experiment groups led us to interpret that the number of training samples and perhaps class balance might be more important for machine learning models than seasonal models (reflecting meteorological and sea ice conditions) at this stage in the move towards sea ice mapping automation, considering the amount of high-resolution benchmark data available. Results do not seem to indicate a consistent decrease in model performance as the scenes in evaluation are farther apart in time from the scenes used for training. However, scenes dominated by water and at the border of experiment groups (February-E, September, October), might have hindered our ability to further investigate this hypothesis. Our models can generally classify water correctly which can bias results positively for scenes with large water polygons. Considering only the main test scenes, January and July, results indicate that using all available data for training generates more reliable models. For ice-water segmentation, the test weighted F1 average (average of January and July 'median' column in Table 5) is 0.96 for *freeze*, 0.97 for *melt*, and 0.98 for *all*, indicating the superior performance when models are trained on year-round data for ice-water discrimination.

Analogous results for oldest ice type are 0.62 for *freeze*, 0.70 for *melt*, and 0.82 for *all*. For dominant ice type: 0.70 for *freeze*, 0.70 for *melt*, and 0.77 for *all*, which again yields the highest performance. These results also show that models trained on *melt* are better than models trained on *freeze*, except for the dominant ice type segmentation where *freeze* and *melt* are tied. We also observed individual *melt* models (#3) achieving best performance for January for the oldest ice type, which indicates how class balance might have a stronger effect than meteorological conditions for deep learning models trained with sufficient samples. *Melt* #3 classified most of January's Old Ice and almost all water correctly for the oldest ice experiment, exactly the classes with higher proportion in the dataset and corresponding to most pixels in the January scene. Having a model that tends to not generate outputs for New Ice, Nilas, and Young Ice improved the overall model performance in that specific case, because Old Ice and Water were classified with higher recall. Having a model trained on the *melt* season achieving better results for type classification on a January scene was unexpected. During melting months, ice is covered with wet snow and melt ponds, which was assumed to be a major drawback for SAR-based interpretation that would be reflected in model performance. However, this might not be a crucial point for machine learning models trained with sufficient samples covering the variety of seasons and ice conditions.

### *5.3. Sentinel-1 noise (sub-swath banding and speckle) and training modes*

Although Stokholm et al. (2022) show minor improvements for sea ice concentration mapping when using improved denoising techniques in their architecture, our results show little to no effect caused by TOPSAR scalloping and banding noise present in the Sentinel-1 EW mode HV polarization. Our models seem to be particularly robust against noise when the encoder is initialized with pre-trained weights. CNNs trained with natural images, such as the ImageNet, tend to learn edge detectors and "blurrers" in the first layers (e.g., Yosinski et al., 2014), a characteristic that tends to remain present after the models are fine-tuned on a secondary task (e.g., Pires de Lima and Duarte, 2021). These Gaussian-like filters (blurrers) seem to account for random speckle noise. The banding noise is directly related to the incidence angle, which we also used as an input channel. The output of algorithms presented in previous research using a patch CNN classification approach (instead of CNN-based semantic segmentation) is heavily impacted by this noise (e.g., Khaleghian et al., 2021), which impacts both ice-water segmentation and type classification. Our segmentation model architecture learns to ignore noise patterns, or put differently, learn the interdependence of noise patterns and SAR backscatter across various incidence angles, leading to high performance in ice-water discrimination. Our interpretation is that the models have an adequate receptive area, allowing them to capture spatial relationships between different regions of the input data. Moreover, with a sufficient number of training examples, the models can learn to ignore systematic and (small) random noise in the data. Although results for models trained with randomly initialized weights might be more susceptible to noise, we do not observe significant noise associated misclassifications. Previous research shows that the size of patches used for training patch classification models seem to play an important role in this aspect as well (e.g., Khaleghian et al., 2021). However, CNN-based semantic segmentation, in contrast with patch classification, does not seem to be as sensitive to TOPSAR-caused noise, and not significantly affected by the size of the patch: we train on sub-patches of the image but predict on full SAR scenes, generating acceptable results especially for ice-water segmentation. The dilation size in atrous convolutions can affect the weights learned during training (e.g., Chen et al., 2017), and further experimentation might highlight potential for improvements with different dilation rates or feature maps dimensions. In general, we also observe only minor misclassifications caused by TOPSAR noise when using the U-Net comparison model.

In general, models that are initialized with pre-trained weights produce better metrics during training and achieve validation plateaus earlier, therefore train faster. Overall, models initialized with pre-trained weights have the potential to generate better metrics for predicting Sentinel-1 scenes as well. More experiments with larger datasets are justified to study the effect of pre-trained weights when training on SAR images.

### 5.4. Future work

Ice charts usually present information on SoD, floe size, as well as partial concentration. Future research could consider different strategies to improve both the performance of the models, as well as their usability. For example, we envision developing a model architecture to predict more than one label using the same feature maps and different decoders, a strategy sometimes called multitask learning. In doing so, the model will be able to predict not only one SoD at a time, but also predict the concentration, secondary SoDs, or any other label associated with the polygons to take maximum advantage of existing information. This is partially inspired by the fact that our dominant ice type label strategy generated slightly better results than the initial oldest ice type strategy. Although we discussed some of the experiment design and hyperparameters choice in this manuscript, including the size of the patch used for training, the choice of encoder, and encoder initialization, the steps taken to improve models' performance is not always clear. We aim to expand in the sensitivity analysis to better help us understand what steps are crucial for improved performance. In order to enhance SoD segmentation and account for the variability of sea ice types, we envision expanding the training datasets and incorporating additional data sources. This includes integrating L-band and multi-frequency SAR data, as the discrimination of certain sea ice types based solely on C-band backscatter values is limited by the distinct dielectric responses exhibited by different ice types towards microwaves. A worthwhile area of focus is generating larger high-resolution and possibly pixel-level benchmark datasets.

## 6. Conclusion

We presented a deep learning model for automating image segmentation for sea ice classification. Our results show high potential to automate the segmentation of sea ice and water using Sentinel-1 backscatter data even when using a small dataset for training. The correct identification of SoD is more challenging, as we observed increased disagreement on the output generated by our models and the data interpreted and labelled by sea ice analysts. Despite the room for improvement with other strategies for SoD characterization, our ice-water models seem to be robust against different seasons and have the potential to be further trained and improved using more labelled data. Our model architecture, which is based on ResNet and Chen et al.'s (2017) ASPP, generates results generally robust against the noise present in Sentinel-1 images. Furthermore, our proposed adaptations in ASPP lead to a model architecture that can efficiently generate outputs for full Sentinel-1 scenes, which reduces issues associated with patch stitching and mosaicing, while accelerating the prediction production for large areas. Our model generates results four times faster (as measured on our hardware), and with a classification performance 2-3% higher when compared with a baseline U-Net. Our proposed architecture is also less affected by the systematic banding noise in Sentinel-1 EW mode and generates predictions with fewer border artifacts. Our results show comparable performance for seasonal and year-round models for SoD segmentation. For ice-water, year-round models have a higher performance. This indicates that increasing the number of available expert-labelled data might yield higher performance improvements than generating seasonal models.

## 7. Acknowledgements

Creation of the Extreme Earth dataset was funded under the European Union's Horizon 2020 research and innovation programme grant agreement 825258 "From Copernicus Big Data to Extreme Earth Analytics". We thank the Extreme Earth project and MET Norway for making this dataset available to the sea ice community at large. This research is supported by the National Science Foundation under Grant No. 2026962. We thank the journal editors as well as three anonymous referees that helped us improve the manuscript during peer-review. The code we developed for this project will be made available at https://github.com/geohai/sea-ice-segment upon publication.

## 8. References

Boulze, Hugo, Anton Korosov, and Julien Brajard. 2020. "Classification of Sea Ice Types in Sentinel-1 SAR Data Using Convolutional Neural Networks." *Remote Sensing* 12 (13). Multidisciplinary Digital Publishing Institute: 2165. doi:10.3390/rs12132165.

Casey, J. Alec, Stephen E. L. Howell, Adrienne Tivy, and Christian Haas. 2016. "Separability of Sea Ice Types from Wide Swath C- and L-Band Synthetic Aperture Radar Imagery Acquired during the Melt Season." *Remote Sensing of Environment* 174 (March): 314–328. doi:10.1016/j.rse.2015.12.021.

Chen, Liang-Chieh, George Papandreou, Iasonas Kokkinos, Kevin Murphy, and Alan L. Yuille. 2015. "Semantic Image Segmentation with Deep Convolutional Nets and Fully Connected CRFs." In *3rd International Conference on Learning Representations, ICLR*, edited by Yoshua Bengio and Yann LeCun. San Diego, CA, USA. http://arxiv.org/abs/1412.7062.

Chen, Liang-Chieh, George Papandreou, Florian Schroff, and Hartwig Adam. 2017. "Rethinking Atrous Convolution for Semantic Image Segmentation." *ArXiv E-Prints*, June, arXiv:1706.05587.

Chen, Xinwei, K. Andrea Scott, Mingzhe Jiang, Yuan Fang, Linlin Xu, and David A. Clausi. 2023. "Sea Ice Classification with Dual-Polarized SAR Imagery: A Hierarchical

Pipeline." In *2023 IEEE/CVF Winter Conference on Applications of Computer Vision Workshops (WACVW)*, 224–232. Waikoloa, HI, USA. doi:10.1109/WACVW58289.2023.00028.

Clausi, David A. 2001. "Comparison and Fusion of Co-occurrence, Gabor and MRF Texture Features for Classification of SAR Sea-ice Imagery." *Atmosphere-Ocean* 39 (3). Taylor & Francis: 183–194. doi:10.1080/07055900.2001.9649675.

Deng, Huawu, and D.A. Clausi. 2005. "Unsupervised Segmentation of Synthetic Aperture Radar Sea Ice Imagery Using a Novel Markov Random Field Model." *IEEE Transactions on Geoscience and Remote Sensing* 43 (3): 528–538. doi:10.1109/TGRS.2004.839589.

Detlefsen, Nicki Skafte, Jiri Borovec, Justus Schock, Ananya Harsh Jha, Teddy Koker, Luca Di Liello, Daniel Stancl, Changsheng Quan, Maxim Grechkin, and William Falcon. 2022. "TorchMetrics - Measuring Reproducibility in PyTorch." *Journal of Open Source Software* 7 (70). The Open Journal: 4101. doi:10.21105/joss.04101.

Dierking, Wolfgang. 2013. "Sea Ice Monitoring by Synthetic Aperture Radar." *Oceanography* 26 (2). Oceanography Society: 100–111.

Durkalec, Agata, Chris Furgal, Mark W. Skinner, and Tom Sheldon. 2015. "Climate Change Influences on Environment as a Determinant of Indigenous Health: Relationships to Place, Sea Ice, and Health in an Inuit Community." *Social Science & Medicine* 136–137 (July): 17–26. doi:10.1016/j.socscimed.2015.04.026.

Falcon, et al., WA. 2019. "PyTorch Lightning." *https://github.com/PyTorchLightning/Pytorch-Lightning*.

Falkingham, J., 2010. SIGRID-3: A Vector Archive Format for Sea Ice Charts. 2010.

Fetterer, F., C. Bertoia, and Jing Ping Ye. 1997. "Multi-Year Ice Concentration from RADARSAT." In , 1:402–404 vol.1. doi:10.1109/IGARSS.1997.615897.

GDAL. 2022. *GDAL/OGR Geospatial Data Abstraction Software Library*. Open Source Geospatial Foundation. doi:10.5281/zenodo.5884351.

Goessling, H. F., S. Tietsche, J. J. Day, E. Hawkins, and T. Jung. 2016. "Predictability of the Arctic Sea Ice Edge." *Geophysical Research Letters* 43 (4): 1642–1650. doi:10.1002/2015GL067232.

Grauman, K., and T. Darrell. 2005. "The Pyramid Match Kernel: Discriminative Classification with Sets of Image Features." In *Tenth IEEE International Conference on Computer Vision (ICCV'05) Volume 1*, 2:1458-1465 Vol. 2. Beijing, China. doi:10.1109/ICCV.2005.239.

Griebel, Jakob, and Wolfgang Dierking. 2018. "Impact of Sea Ice Drift Retrieval Errors, Discretization and Grid Type on Calculations of Ice Deformation." *Remote Sensing* 10 (3). Multidisciplinary Digital Publishing Institute: 393. doi:10.3390/rs10030393.

Gupta, Mukesh. 2015. "Various Remote Sensing Approaches to Understanding Roughness in the Marginal Ice Zone." *Physics and Chemistry of the Earth, Parts A/B/C*, Emerging science and applications with microwave remote sensing data, 83–84 (January): 75–83. doi:10.1016/j.pce.2015.05.003.

Haverkamp, D., Leen Kiat Soh, and C. Tsatsoulis. 1995. "A Comprehensive, Automated Approach to Determining Sea Ice Thickness from SAR Data." *IEEE Transactions on Geoscience and Remote Sensing* 33 (1): 46–57. doi:10.1109/36.368223.

He, Kaiming, Xiangyu Zhang, Shaoqing Ren, and Jian Sun. 2014. "Spatial Pyramid Pooling in Deep Convolutional Networks for Visual Recognition." In *Computer Vision – ECCV 2014*, edited by David Fleet, Tomas Pajdla, Bernt Schiele, and Tinne Tuytelaars, 346–361. Lecture Notes in Computer Science. Cham, Switzerland: Springer International Publishing. doi:10.1007/978-3-319-10578-9_23.

He, Kaiming, Xiangyu Zhang, Shaoqing Ren, and Jian Sun. 2016. "Deep Residual Learning for Image Recognition." In *2016 IEEE Conference on Computer Vision and Pattern Recognition (CVPR)*, 770–778. IEEE. doi:10.1109/CVPR.2016.90.

Ho, Joshua. 2010. "The Implications of Arctic Sea Ice Decline on Shipping." *Marine Policy* 34 (3): 713–715. doi:10.1016/j.marpol.2009.10.009.

Holschneider, M., R. Kronland-Martinet, J. Morlet, and Ph. Tchamitchian. 1990. "A Real-Time Algorithm for Signal Analysis with the Help of the Wavelet Transform." In *Wavelets*, edited by Jean-Michel Combes, Alexander Grossmann, and Philippe Tchamitchian, 286–297. Inverse Problems and Theoretical Imaging. Berlin, Heidelberg: Springer. doi:10.1007/978-3-642-75988-8_28.

Holt, Benjamin, Ron Kwok, and E. Rignot. 1989. "Ice Classification Algorithm Development and Verification for the Alaska Sar Facility Using Aircraft Imagery." In *12th Canadian Symposium on Remote Sensing Geoscience and Remote Sensing Symposium*, 2:751–754.

Howell, Stephen E. L., Alexander S. Komarov, Mohammed Dabboor, Benoit Montpetit, Michael Brady, Randall K. Scharien, Mallik S. Mahmud, Vishnu Nandan, Torsten Geldsetzer, and John J. Yackel. 2018. "Comparing L- and C-Band Synthetic Aperture Radar Estimates of Sea Ice Motion over Different Ice Regimes." *Remote Sensing of Environment* 204 (January): 380–391. doi:10.1016/j.rse.2017.10.017.

Hughes, Nick, and Frank Amdal. 2021. "ExtremeEarth Polar Use Case Training Data." Zenodo. doi:10.5281/zenodo.4683174.

JCOMM Expert Team on Sea Ice, 2014. SIGRID-3 : A vector archive format for sea ice charts: developed by the International Ice Charting Working Group's Ad Hoc Format Team for the WMO Global Digital Sea Ice Data Bank Project. Geneva, Switzerland, WMO & IOC, 24pp. (WMO TD: 1214), (JCOMM Technical Report, 23). http://dx.doi.org/10.25607/OBP-1498

JCOMM Expert Team on Sea Ice, 2017. Sea ice information services of the world, Edition 2017. Geneva, Switzerland, World Meteorological Organization, 103pp. (WMO-No.574). DOI: http://dx.doi.org/10.25607/OBP-1325

Karvonen, Juha. 2017. "Baltic Sea Ice Concentration Estimation Using SENTINEL-1 SAR and AMSR2 Microwave Radiometer Data." *IEEE Transactions on Geoscience and Remote Sensing* 55 (5): 2871–2883. doi:10.1109/TGRS.2017.2655567.

Khaleghian, Salman, Habib Ullah, Thomas Kræmer, Nick Hughes, Torbjørn Eltoft, and Andrea Marinoni. 2021. "Sea Ice Classification of SAR Imagery Based on Convolution Neural

Networks." *Remote Sensing* 13 (9). Multidisciplinary Digital Publishing Institute: 1734. doi:10.3390/rs13091734.

Kingma, Diederik P, and Jimmy Ba. 2014. "Adam: A Method for Stochastic Optimization." In *ICLR*, arXiv:1412.6980. San Diego, CA, USA: ICLR.

Lazebnik, S., C. Schmid, and J. Ponce. 2006. "Beyond Bags of Features: Spatial Pyramid Matching for Recognizing Natural Scene Categories." In *2013 IEEE Conference on Computer Vision and Pattern Recognition*, 2:2169–2178. Los Alamitos, CA, USA: IEEE Computer Society. doi:10.1109/CVPR.2006.68.

Lecun, Y., L. Bottou, Y. Bengio, and P. Haffner. 1998. "Gradient-Based Learning Applied to Document Recognition." *Proceedings of the IEEE* 86 (11): 2278–2324. doi:10.1109/5.726791.

Leppäranta, M., 2009. Sea Ice Dynamics, in: Steele, J.H. (Ed.), Encyclopedia of Ocean Sciences (Second Edition). Academic Press, Oxford, pp. 159–169. https://doi.org/10.1016/B978-012374473-9.00640-8

Lohse, Johannes, Anthony P. Doulgeris, and Wolfgang Dierking. 2020. "Mapping Sea-Ice Types from Sentinel-1 Considering the Surface-Type Dependent Effect of Incidence Angle." *Annals of Glaciology* 61 (83). Cambridge University Press: 260–270. doi:10.1017/aog.2020.45.

Lohse, Johannes, Anthony P. Doulgeris, and Wolfgang Dierking. 2021. "Incident Angle Dependence of Sentinel-1 Texture Features for Sea Ice Classification." *Remote Sensing* 13 (4). doi:10.3390/rs13040552.

McHugh, Mary L. 2012. "Interrater Reliability: The Kappa Statistic." *Biochemia Medica* 22 (3). Croatian Society of Medical Biochemistry and Laboratory Medicine: 276–282.

Meier, Walter N., Greta K. Hovelsrud, Bob E.H. van Oort, Jeffrey R. Key, Kit M. Kovacs, Christine Michel, Christian Haas, et al. 2014. "Arctic Sea Ice in Transformation: A Review of Recent Observed Changes and Impacts on Biology and Human Activity." *Reviews of Geophysics* 52 (3): 185–217. doi:10.1002/2013RG000431.

Melia, N., K. Haines, and E. Hawkins. 2016. "Sea Ice Decline and 21st Century Trans-Arctic Shipping Routes." *Geophysical Research Letters* 43 (18): 9720–9728. doi:10.1002/2016GL069315.

Olah, Chris, Alexander Mordvintsev, and Ludwig Schubert. 2017. "Feature Visualization." *Distill*. doi:10.23915/distill.00007.

Park, Jeong-Won, Anton Andreevich Korosov, Mohamed Babiker, Joong-Sun Won, Morten Wergeland Hansen, and Hyun-Cheol Kim. 2020. "Classification of Sea Ice Types in Sentinel-1 Synthetic Aperture Radar Images." *The Cryosphere* 14 (8). Copernicus GmbH: 2629–2645. doi:10.5194/tc-14-2629-2020.

Park, Jeong-Won, Joong-Sun Won, Anton A. Korosov, Mohamed Babiker, and Nuno Miranda. 2019. "Textural Noise Correction for Sentinel-1 TOPSAR Cross-Polarization Channel Images." *IEEE Transactions on Geoscience and Remote Sensing* 57 (6): 4040–4049. doi:10.1109/TGRS.2018.2889381.

Paszke, Adam, Sam Gross, Francisco Massa, Adam Lerer, James Bradbury, Gregory Chanan, Trevor Killeen, et al. 2019. "PyTorch: An Imperative Style, High-Performance Deep Learning Library." In *Advances in Neural Information Processing Systems 32*, edited by H. Wallach, H. Larochelle, A. Beygelzimer, F. Alché-Buc, E. Fox, and R. Garnett, 8024–8035. http://papers.neurips.cc/paper/9015-pytorch-an-imperative-style-high-performance-deep-learning-library.pdf.

Pedregosa, F, G Varoquaux, A Gramfort, V Michel, B Thirion, O Grisel, M Blondel, et al. 2011. "Scikit-Learn: Machine Learning in Python." *Journal of Machine Learning Research* 12: 2825–2830.

Pires de Lima, Rafael, and David Duarte. 2021. "Pretraining Convolutional Neural Networks for Mudstone Petrographic Thin-Section Image Classification." *Geosciences* 11 (8). Multidisciplinary Digital Publishing Institute: 336. doi:10.3390/geosciences11080336.

Pizzolato, Larissa, Stephen E. L. Howell, Jackie Dawson, Frédéric Laliberté, and Luke Copland. 2016. "The Influence of Declining Sea Ice on Shipping Activity in the Canadian Arctic." *Geophysical Research Letters* 43 (23): 12,146-12,154. doi:10.1002/2016GL071489.

Raney, R., and J. Falkingham. 1994. "RADARSAT and Operational Ice Information." In *Remote Sensing of Sea Ice and Icebergs*, 686. Wiley Series in Remote Sensing and Image Processing. Wiley.

Ren, Yibin, Xiaofeng Li, Xiaofeng Yang, and Huan Xu. 2021. "Development of a Dual-Attention U-Net Model for Sea Ice and Open Water Classification on SAR Images." *IEEE Geoscience and Remote Sensing Letters* 19: 1–5. doi:10.1109/LGRS.2021.3058049.

Ressel, Rudolf, Anja Frost, and Susanne Lehner. 2015. "A Neural Network-Based Classification for Sea Ice Types on X-Band SAR Images." *IEEE Journal of Selected Topics in Applied Earth Observations and Remote Sensing* 8 (7): 3672–3680. doi:10.1109/JSTARS.2015.2436993.

Ronneberger, Olaf, Philipp Fischer, and Thomas Brox. 2015. "U-Net: Convolutional Networks for Biomedical Image Segmentation." In *Medical Image Computing and Computer-Assisted Intervention – MICCAI 2015*, edited by Nassir Navab, Joachim Hornegger, William M Wells, and Alejandro F Frangi, 234–241. Cham: Springer International Publishing.

Russakovsky, Olga, Jia Deng, Hao Su, Jonathan Krause, Sanjeev Satheesh, Sean Ma, Zhiheng Huang, et al. 2015. "ImageNet Large Scale Visual Recognition Challenge." *International Journal of Computer Vision* 115 (3). Springer US: 211–252. doi:10.1007/s11263-015-0816-y.

Saldo, Roberto, Matilde Brandt Kreiner, Jørgen Buus-Hinkler, Leif Toudal Pedersen, David Malmgren-Hansen, Allan Aasbjerg Nielsen, and Henning Skriver. 2021. "AI4Arctic / ASIP Sea Ice Dataset - Version 2," June. doi:10.11583/DTU.13011134.v3.

Singha, Suman, A. Malin Johansson, and Anthony P. Doulgeris. 2021. "Robustness of SAR Sea Ice Type Classification Across Incidence Angles and Seasons at L-Band." *IEEE Transactions on Geoscience and Remote Sensing* 59 (12): 9941–9952. doi:10.1109/TGRS.2020.3035029.

Stokholm, Andreas, Tore Wulf, Andrzej Kucik, Roberto Saldo, Jørgen Buus-Hinkler, and Sine Munk Hvidegaard. 2022. "AI4SeaIce: Toward Solving Ambiguous SAR Textures in Convolutional Neural Networks for Automatic Sea Ice Concentration Charting." *IEEE Transactions on Geoscience and Remote Sensing* 60: 1–13. doi:10.1109/TGRS.2022.3149323.

Wang, Lei, K. Andrea Scott, and David A. Clausi. 2017. "Sea Ice Concentration Estimation during Freeze-Up from SAR Imagery Using a Convolutional Neural Network." *Remote Sensing* 9 (5). doi:10.3390/rs9050408.

Wang, Lei, K. Andrea Scott, Linlin Xu, and David A. Clausi. 2016. "Sea Ice Concentration Estimation During Melt From Dual-Pol SAR Scenes Using Deep Convolutional Neural Networks: A Case Study." *IEEE Transactions on Geoscience and Remote Sensing* 54 (8): 4524–4533. doi:10.1109/TGRS.2016.2543660.

Wang, Yi-Ran, and Xiao-Ming Li. 2021. "Arctic Sea Ice Cover Data from Spaceborne Synthetic Aperture Radar by Deep Learning." *Earth System Science Data* 13 (6). Copernicus GmbH: 2723–2742. doi:10.5194/essd-13-2723-2021.

Yosinski, Jason, Jeff Clune, Yoshua Bengio, and Hod Lipson. 2014. "How Transferable Are Features in Deep Neural Networks?" *Advances in Neural Information Processing Systems* 27 (November): 3320–3328.

Yu, Fisher, and Vladlen Koltun. 2016. "Multi-Scale Context Aggregation by Dilated Convolutions." In *4th International Conference on Learning Representations, ICLR*, edited by Yoshua Bengio and Yann LeCun. San Juan, Puerto Rico. http://arxiv.org/abs/1511.07122.

Zakhvatkina, Natalia, Anton Korosov, Stefan Muckenhuber, Stein Sandven, and Mohamed Babiker. 2017. "Operational Algorithm for Ice–Water Classification on Dual-Polarized RADARSAT-2 Images." *The Cryosphere* 11 (1). Copernicus GmbH: 33–46. doi:10.5194/tc-11-33-2017.

Zakhvatkina, Natalia, Vladimir Smirnov, and Irina Bychkova. 2019. "Satellite SAR Data-Based Sea Ice Classification: An Overview." *Geosciences* 9 (4). Multidisciplinary Digital Publishing Institute: 152. doi:10.3390/geosciences9040152.

Zeiler, Matthew D., and Rob Fergus. 2014. "Visualizing and Understanding Convolutional Networks." In *Computer Vision – ECCV 2014*, edited by David Fleet, Tomas Pajdla, Bernt Schiele, and Tinne Tuytelaars, 818–833. Lecture Notes in Computer Science. Cham: Springer International Publishing. doi:10.1007/978-3-319-10590-1_53.